\documentclass[journal]{IEEEtran}
\usepackage{graphicx}
\usepackage{epstopdf}
\usepackage{fancyhdr}
\usepackage{amsmath,amssymb,amstext}
\usepackage{subeqnarray}
\usepackage{cite}
\usepackage{hhline}
\usepackage{bm}
\usepackage{soul}
\usepackage{slashbox}
\usepackage{times,amsmath,amsfonts}
\usepackage{subfigure}
\usepackage{amsmath,amssymb}
\usepackage{graphicx,picture}
\usepackage{cite}
\usepackage{subfigure}
\usepackage{subeqnarray}
\usepackage{cases}
\usepackage{color}
\usepackage{overpic}
\usepackage{multirow}
\usepackage{booktabs}
\usepackage{hyperref}
\ifCLASSINFOpdf
\else
   \usepackage{graphicx}
   \graphicspath{{../eps/}}
   \DeclareGraphicsExtensions{.eps}
\fi
\usepackage{fancyhdr}
\usepackage{amsmath,amssymb,amstext}
\usepackage{subeqnarray}
\usepackage{cite}
\usepackage{hhline}
\usepackage{bm}
\usepackage{soul}
\usepackage{slashbox}
\usepackage{times,amsmath,amsfonts}
\usepackage{subfigure}
\usepackage{amsmath,amssymb}
\usepackage{graphicx,picture}
\usepackage{cite}
\usepackage{subfigure}
\usepackage{subeqnarray}
\usepackage{cases}
\usepackage{color}
\usepackage{overpic}
\usepackage{multirow}
\usepackage{booktabs}
\usepackage{epstopdf}

\definecolor{r}{rgb}{1,0,0}
\definecolor{b}{rgb}{0,0,1}
\definecolor{k}{rgb}{0,1,1}
\usepackage{upgreek}

\newcounter{saveeqn}%

\allowdisplaybreaks
\DeclareMathSymbol{\Phi}{\mathord}{letters}{8}

\begin{document}
\title{Waveform Design and Performance Analysis for Full-Duplex Integrated Sensing and Communication}

\author{
\IEEEauthorblockN{Zhiqiang Xiao,~\IEEEmembership{Student Member, IEEE} and Yong Zeng,~\IEEEmembership{Member, IEEE}}

\thanks{
The authors are with the National Mobile Communications Research Laboratory, Southeast University, Nanjing 210096, China. Y. Zeng is also with the Purple Mountain Laboratories, Nanjing 211111, China (e-mail: \{zhiqiang\_xiao, yong\_zeng\}@seu.edu.cn). (\emph{Corresponding author: Yong Zeng.})

This work was supported by the National Key R\&D Program of China with Grant number 2019YFB1803400.
Part of this work has been submitted to the IEEE GLOBECOM 2021, Madrid, Spain, 7-11 Dec. 2021 \cite{full2021xiao}.
}
}
\maketitle

\begin{abstract}
Integrated sensing and communication (ISAC) is a promising technology to fully utilize the precious spectrum and hardware in wireless systems, which has attracted significant attentions recently.
This paper studies ISAC for the important and challenging monostatic setup, where one single ISAC node wishes to simultaneously sense a radar target while communicating with a communication receiver.
Different from most existing schemes that rely on either radar-centric half-duplex (HD) pulsed transmission with information embedding that suffers from extremely low communication rate, or communication-centric waveform that suffers from degraded sensing performance, we propose a novel full-duplex (FD) ISAC scheme that utilizes the waiting time of conventional pulsed radars to transmit dedicated communication signals.
Compared to radar-centric pulsed waveform with information embedding, the proposed design can drastically increase the communication rate, and also mitigate the sensing eclipsing and near-target blind range issues, as long as the self-interference (SI) is effectively suppressed.
On the other hand, compared to communication-centric ISAC waveform, the proposed design has better auto-correlation property as it preserves the classic  radar waveform for sensing.
Performance analysis is developed by taking into account the residual SI, in terms of the probability of detection and ambiguity function for sensing, as well as the spectrum efficiency for communication.
Numerical results are provided to show the significant performance gain of our proposed design over benchmark schemes.
\end{abstract}


\section{Introduction}
Wireless communication and radar sensing, as two most successful applications of electromagnetic radiation, were mainly developed separately in the past century, with quite different performance metrics and distinct frequency bands \cite{richards2010principles,TS38.101}.
With the continuous expansion of communication spectrum and the growing interest to bridge the cyber and physical worlds by ubiquitous sensing, there have been growing research interest in \emph{integrated sensing and communication} (ISAC) \cite{paul2016survey,mishra2019toward,liu2020joint,liu2018mu,tan2021integrated,liu2021survey,zhang2021overview,yang2021integrated,wang2021snr}, which aims to efficiently utilize the precious radio resources and hardware for both sensing and communication purposes.

The concept related to ISAC can be traced back to 1960s~\cite{mealey1963method}, where pulse code groups were transmitted for radar sensing, with some pulses in each group used to carry information.
However, due to the significant difference in hardware components and signal processing procedures between conventional radar sensing and wireless communication, the research on the joint consideration of radar and communication was rare in the subsequent decades.
With the advancement of modern radar and wireless communication systems, the hardware and signal processing procedures for both systems become more similar, which brings the opportunities to pursue ISAC for more efficient use of hardware, spectrum and energy \cite{paul2016survey,mishra2019toward,liu2020joint}.
Furthermore, with the ever-increasing requirement on wireless communication rate, the frequency bands that were usually considered for radar (e.g., {\it L} and {\it S}-bands \cite{richards2010principles}) are also used by communication systems, such as the 5G new radio (NR)~\cite{TS38.101,liu2020joint}, which renders ISAC necessary to improve the performance of both sensing and communication.
In fact, ISAC makes it possible to even achieve mutualism between radio sensing and communication.
For example, the sensed information like the angle, range, and location of the user equipment (UE) or even scatterers can be used for communication performance enhancement, such as sensing-assisted beamforming \cite{liu2020radar,ali2020leveraging} and environment-aware resource allocation and beam alignment \cite{zeng2021toward,wu2021environment}.
On the other hand, the contemporary cellular communication network, which is almost ubiquitously available with well-established infrastructure and powerful signal processing capability, is expected to facilitate the realization of ubiquitous sensing to bridge the cyber and physical worlds in the future~\cite{saad2019vision,zhang2020perceptive,xiao2020overview}.

Earlier research efforts on ISAC-related study were mainly devoted to {\it radar-communication coexistence} (RCC) \cite{paul2016survey,zheng2019radar}, where the radar and communication systems are separately designed and treat each other as the detrimental interference that needs to be appropriately mitigated.
To this end, various interference mitigation techniques between radar and communication systems have been proposed, such as those based on opportunistic spectrum sharing \cite{saruthirathanaworakun2012opportunistic,cohen2017spectrum}, or null space projection (NSP) using multiple antennas \cite{sodagari2012projection,khawar2015target}.

Compared to RCC, the more aggressive vision is to seamlessly integrate radar sensing and wireless communication into a common device with small form factor, which is also known as {\it dual-function radar-communication} (DFRC) \cite{blunt2010intrapulse,sturm2011waveform,hassanien2016signaling}.
Significant research efforts have been devoted to the waveform design and performance analysis of DFRC systems, which can be loosely classified as radar-centric \cite{hassanien2015dual,huang2020majorcom,nowak2016co} or communication-centric \cite{braun2014ofdm,gaudio2020effectiveness}.
For radar-centric approaches, the communication symbols are usually embedded into the radar waveforms \cite{hassanien2016signaling}, which typically have extremely low communication rate.
Typical information-embedding methods include beampattern modulation \cite{hassanien2015dual}, index modulation \cite{huang2020majorcom}, and fast-time modulation \cite{nowak2016co}.
For beampattern modulation \cite{hassanien2015dual}, information embedding is achieved by controlling the sidelobe of the radar beampattern.
However, such methods require that the mainlobe of the radar beampattern points to the target, which restricts the flexibility for communication.
For index modulation \cite{huang2020majorcom}, information is embedded based on the index of certain radio resources that have active states, such as the index of the antenna or the carrier frequency used for transmission.
For fast-time modulation \cite{nowak2016co}, communication symbols are embedded over the fast-time of radar waveforms, which can increase the communication rate as compared to the above two methods but at the cost of compromised radar performance.

On the other hand, for communication-centric ISAC \cite{braun2014ofdm,gaudio2020effectiveness}, radar sensing is achieved by directly using communication waveforms, such as orthogonal frequency division multiplexing (OFDM) waveform \cite{sturm2011waveform,braun2014ofdm}. Another emerging modulation technique termed {\it orthogonal time frequency space} (OTFS) modulation \cite{hadani2017orthogonal} has also been studied for ISAC \cite{gaudio2020effectiveness}, which is more robust to large Doppler frequency shifts than OFDM.
However, communication waveforms are random in nature, which usually lead to degraded sensing performance due to various issues like high peak-to-average-power ratio (PAPR), random autocorrelation property, and compromised resolution~\cite{sturm2011waveform}.
To address such issues, there are some research efforts towards to the joint radar-communication (JRC) design \cite{yuan2020spatio,liu2020joint2,liu2017multiobjective,kumari2019adaptive}, where the performance metrics for both communication and sensing are jointly considered, such as those based on mutual information (MI) \cite{yuan2020spatio}, waveform similarity \cite{liu2020joint2}, Cramer-Rao lower bound (CRLB) \cite{liu2017multiobjective}, and joint coding \cite{kumari2019adaptive}.
However, such joint design methods usually involve sophisticated optimization problems, which are difficult and time-consuming to solve.
Besides, those performance metrics like MI and waveform similarity do not directly reflect the sensing performance, and the CRLB-based joint performance optimization is quite challenging to solve due to the complicated expressions of CRLBs.

In this paper, we consider the ISAC system for the important and challenging monostatic setup \cite{barneto2021full}, where one single ISAC node wishes to simultaneously sense a radar target while communicating with a communication receiver.
Different from most existing monostatic ISAC schemes that rely on either half-duplex (HD) pulse transmission with information embedding, which suffers from extremely low communication rate, or communication-centric waveform that suffers from degraded sensing performance,
we propose a novel full-duplex (FD) ISAC scheme that fully utilizes the waiting time in the conventional pulsed radar to transmit dedicated communication signals, so as to achieve high sensing and communication performance.
The main contributions of this paper are summarized as follows.
\begin{itemize}
\item
First, we propose a low-complexity waveform design for the challenging monostatic ISAC system, which exploits the FD capability of the ISAC node to time-multiplexing the dedicated sensing and communication waveforms.
The proposed design does not require any complicated optimization and it includes the conventional pulsed radar waveform and continuous communication waveforms as special cases by appropriate power control.
Compared to radar-centric pulsed waveform with information embedding, the proposed design can drastically increase the communication spectrum efficiency, and also mitigate the eclipsing and near-target blind range issues, as long as the self-interference (SI) is effectively suppressed.
On the other hand, compared to communication-centric waveform, the proposed design has better autocorrelation property as it preserves the classic radar waveform for sensing.
\item
Next, rigorous performance analysis is performed for the proposed design, by taking into account the residual SI.
Specifically, the probability of detection and the ambiguity function (AF) are derived for sensing, and the symbol error rate and spectrum efficiency are derived for communication.
It is revealed that depending on the SI cancelation (SIC) capability, the probability of detection of our proposed FD-ISAC scheme can be even higher than that of the conventional pulsed radar, thanks to the additional sensing energy provided by the dedicated communication signals.
\item
Finally, extensive numerical results are provided to compare the performance of the proposed FD-ISAC scheme with various benchmark waveforms, in terms of the communication rate, probability of detection, maximum detectable range, autocorrelation function (ACF), which demonstrate the great potential of our proposed FD-ISAC scheme.
\end{itemize}

The rest of this paper is organized as follows. Section~\ref{system_model} introduces the model of the monostatic FD-ISAC system.
After a brief review on the classic pulsed radar waveform with information embedding, we present the proposed FD-ISAC waveform design.
In Section~\ref{sensing_performance}, the sensing performance of the proposed scheme in terms of the probability of detection and the AF is analyzed.
In Section~\ref{communication_performance}, the communication performance of the proposed design in terms of symbol error rate and spectrum efficiency is analyzed.
In Section~\ref{numerical_results}, numerical results are provided to evaluate the performance of our proposed design and finally, we conclude the paper in Section~\ref{conclusion}.

\section{System Model and Proposed Full-Duplex ISAC Waveform}\label{system_model}
\begin{figure} 
  \centering
  \subfigure[Collocated sensing target and communication receiver]{
  \includegraphics[width=0.48\textwidth]{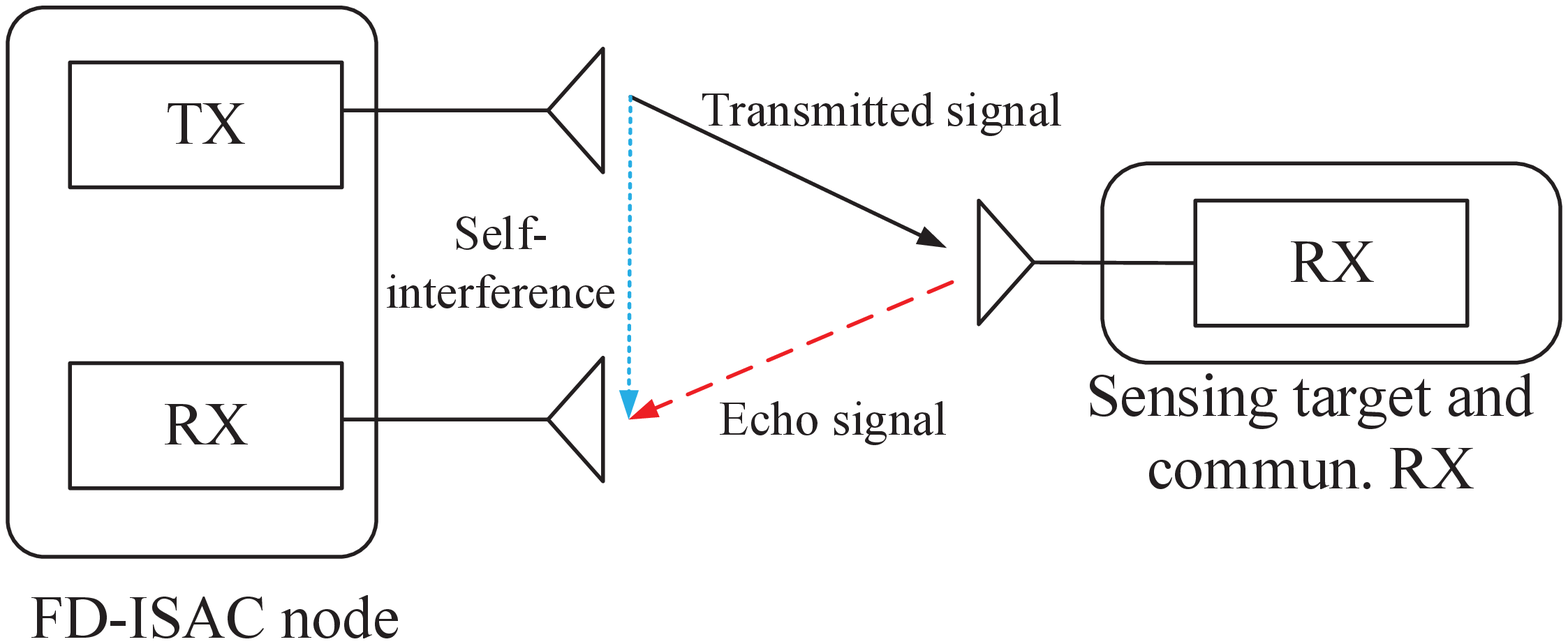}\label{collocated_system}
  }
  \subfigure[Separated sensing target and communication receiver]{
  \includegraphics[width=0.48\textwidth]{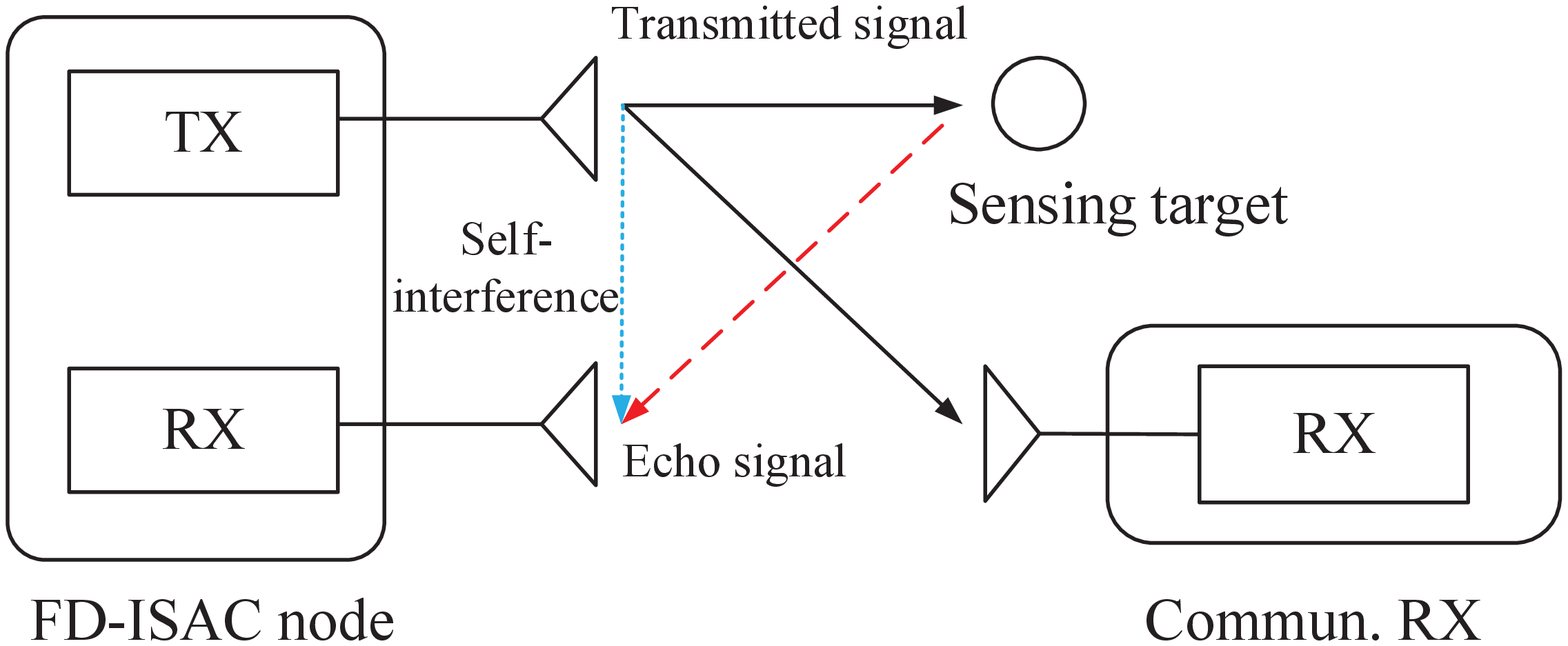}\label{seperated_system}
  }
  \caption{Two typical architectures for FD-ISAC system: (a) collocated sensing target and communication receiver; (b) separated sensing target and communication receiver.}\label{system}\vspace{-10pt}
\end{figure}

As illustrated in Fig.~\ref{system}, we consider a monostatic ISAC system, where one single FD-ISAC node wishes to simultaneously sense a radar target while communicating with a communication receiver.
To this end, FD-ISAC node is equipped with separated transmit and receive antennas, for simultaneous signal transmission and radar echo reception, respectively.
Furthermore, depending on whether the communication receiver and sensing target are collocated or separated, we have the collocated and separated FD-ISAC architectures, as shown in Fig.~\ref{collocated_system} and Fig.~\ref{seperated_system}, respectively.
In the collocated architecture, the FD-ISAC node needs to communicate with a target and also sense its state, such as its distance and moving velocity.
One typical application scenario is unmanned aerial vehicle (UAV) swarm operation \cite{zeng2019accessing}, where the FD-ISAC node corresponds to a UAV cluster head and the collocated communication receiver/sensing target corresponds to the UAV follower.
On the other hand, for FD-ISAC with separated architecture, the FD-ISAC node needs to sense a target while communicating with a different communication receiver.
One typical application scenario is intelligent vehicular-to-everything (V2X) network, where the FD-ISAC node, sensing target, and communication receiver could correspond to the roadside unit (RSU), surrounding vehicle, and pedestrian, respectively.
In the following, the main techniques and results are presented based on the separate architecture, while they can also be applied to the collocated architecture in a straightforward manner.

Let $B$ be the total bandwidth available for the ISAC system.
The general baseband waveform transmitted by the FD-ISAC node over one radar {\it coherent processing interval} (CPI) can be expressed as
\begin{equation}\label{x_t}
x(t) = \sum\limits_{k=0}^{K-1} x_k(t-kT), 0\le t\le KT,
\end{equation}
where $T\gg \frac{1}{B}$ is the {\it pulse repetition interval} (PRI), $K$ is the number of PRIs for each CPI, and $x_k(t)$ is the waveform sent over the $k$-th PRI, which has the bandwidth $B$.

In the following, before presenting our proposed waveform design for $x_k(t)$ in each PRI, we first give a brief overview on the conventional pulsed waveform that was designed for radar sensing only, as well as the radar-centric ISAC waveform with low-rate information symbol embedding.

\subsection{Conventional Pulsed Waveform}\label{Conventional radar}
\begin{figure} 
  \centering
  \subfigure[Conventional HD pulsed radar waveform]{
  \includegraphics[width=0.45\textwidth]{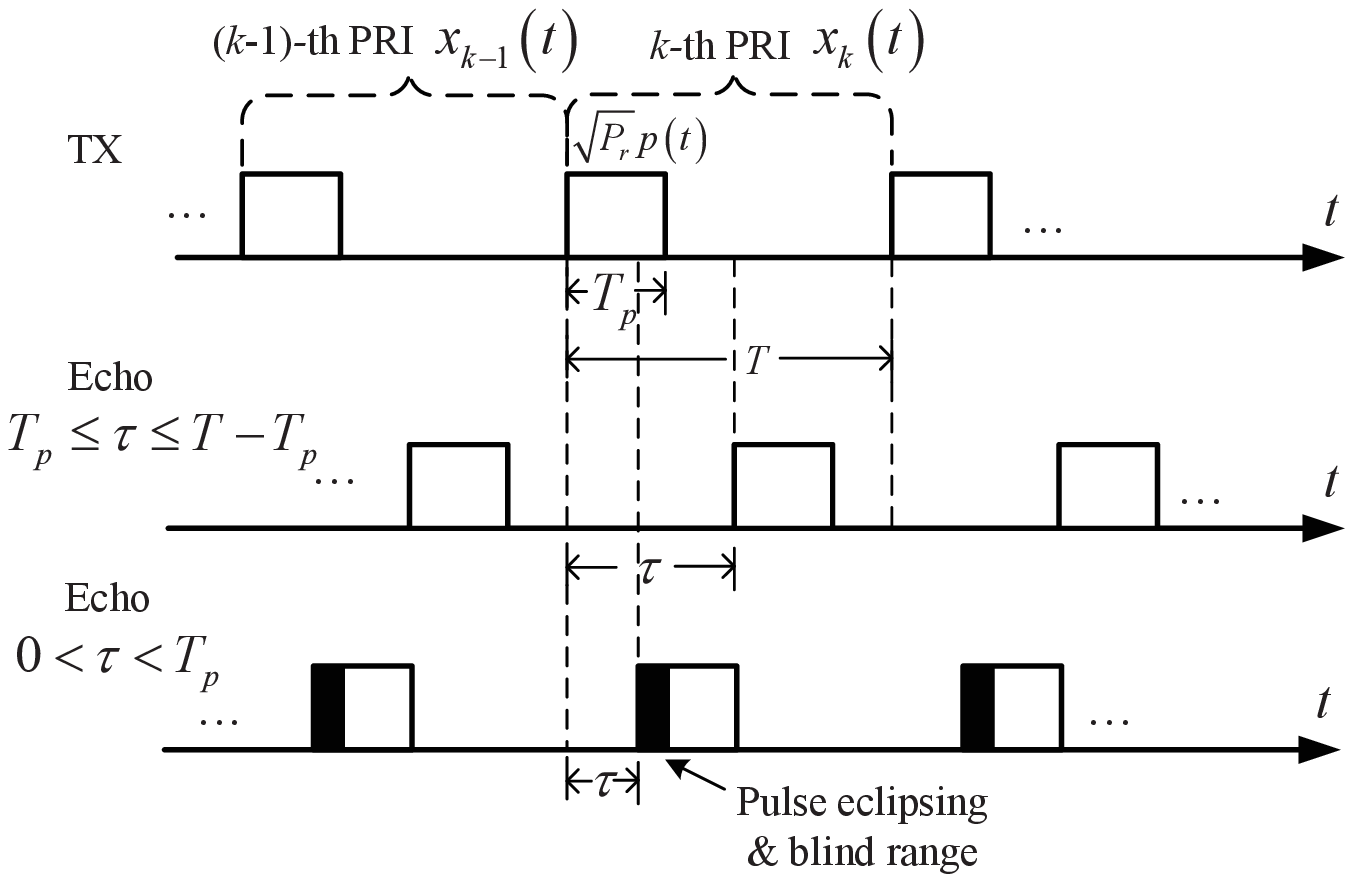}\label{hd}
  }
  \subfigure[Proposed FD-ISAC waveform]{
  \includegraphics[width=0.45\textwidth]{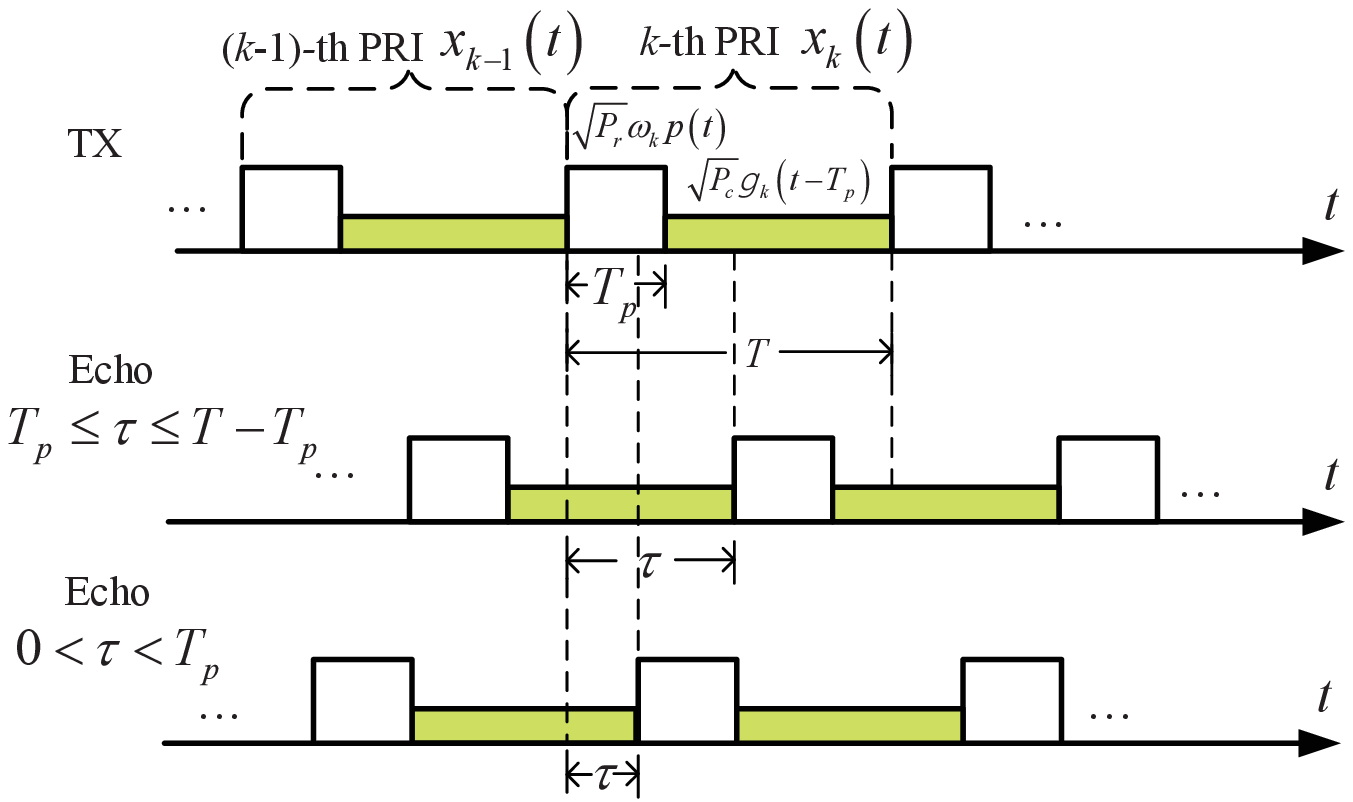}\label{fd}
  }\\
  \caption{The conventional HD pulsed radar waveform versus the proposed FD-ISAC waveform.}\vspace{-10pt}
\end{figure}

As illustrated in Fig.~\ref{hd}, a classic waveform for monostatic radar sensing is the HD pulsed waveform, where the monostatic radar alternates between pulse transmission and echo reception, and the radar receiver is usually switched off while transmitting so as to avoid SI \cite{richards2010principles}.
Let $T_p<T$ denote the radar pulse duration.
Then the transmitted signal $x_k(t)$ for the $k$-th PRI in \eqref{x_t} can be written as
\begin{equation}\label{x_k}
x_k(t)=\left\{\begin{split}
&\sqrt{P_r}p(t),&&0\le t\le T_p  \\
&0,&&T_p< t \le T
\end{split}
\right.,
k=0,\cdots,K-1,
\end{equation}
where $P_r$ is the radar transmit power; $p(t)$ is the basic radar pulse with bandwidth $B$ and normalized power, i.e., $\frac{1}{T_p}\int_0^{T_p}|p(t)|^2 \mathrm{d}t=1$.
The ratio $\rho=T_p/T$ is known as the {\it duty cycle}.
Note that the classic  continuous waveform (CW) radar waveform, such as frequency-modulated CW (FMCW), can be viewed as a special case of \eqref{x_k} with $\rho=1$, but that will incur severe SI \cite{blunt2016overview}.
To mitigate the SI, the conventional CW radar needs to limit its transmit power, which thus greatly restricts its sensing range and is only suitable for nearby target sensing \cite{richards2010principles}.
On the other hand, for typical pulsed radar waveforms, the underlying assumption is that the pulse duration is sufficiently short, i.e., $T_p\ll T$ (or $\rho\ll 1$), and the target is sufficiently far away from the radar receiver \cite{blunt2016overview}, so that for each PRI, the target echo only arrives after each pulse transmission is completed, i.e., $T_p\le\tau\le T-T_p$, where $\tau$ is two-way propagation delay.
However, as illustrated in Fig.~\ref{hd}, for nearby target sensing where the target echo returns while the radar pulse is still transmitting, a portion of the echo cannot be captured by the radar receiver, which degrades the received signal-to-noise ratio (SNR) and range resolution, an issue known as {\it pulse eclipsing} \cite{blunt2016overview}.
In general, the minimal measurable range of the pulsed radar is $R_{\min}=\frac{c(T_p+t_{r})}{2}$, where $t_r$ is the recovery time of the radar system for switching from transmission to reception, and $c$ is the speed of light.
Therefore, the target at a range smaller than $R_{\min}$ cannot be detected, which is also referred to as the {\it blind range} \cite{richards2010principles}.
For example, a pulse duration with $T_p=1$ microsecond ($\upmu s$) corresponds to the blind range about $150$ meters (m), which is acceptable for conventional long-range radar applications, but is unsuitable for future ISAC applications like UAV swarm or V2X networks with both distant and nearby targets.

For the basic radar pulse $p(t)$, if a simple unmodulated pulse is used, the pulse duration $T_p$ and bandwidth $B$ are related as $T_p=1/B$, or the time-bandwidth product is $BT_p=1$.
However, the range resolution (i.e., $\triangle R\approx \frac{c}{2B}$) is inversely proportional to $B$ while the pulse energy is proportional to $T_p$.
Therefore, for the simple unmodulated radar waveform where $BT_p=1$, it is well known that there exists a severe compromise between improving the radar resolution and energy \cite{richards2010principles}.
Such an issue can be resolved by applying the pulse compression technique \cite{richards2010principles}, where the radar pulse $p(t)$ is appropriately modulated so that the time-bandwidth product is much greater than 1, i.e., $N=BT_p\gg 1$.
Typical pulse compression waveforms include linear frequency modulation (LFM) and phase-coded waveforms \cite{blunt2016overview}.

With pulse compression, the basic radar pulse with bandwidth $B$ and time-bandwidth product $N$ (or pulse duration $T_p=N/B$) can be expressed as \cite{zheng2019radar}
\begin{equation}\label{p_t}
p(t) = \sqrt{T_c}\sum\limits_{n=0}^{N-1}c[n]\psi(t-nT_c), 0\le t\le T_p,
\end{equation}
where $T_c=\frac{T_p}{N}=\frac{1}{B}$ is the chip duration;
$\psi(t)$ is the Nyquist waveform of bandwidth $B=1/T_c$, whose autocorrelation $R_{\psi}(t)$ satisfies $R_{\psi}(t)=\delta(t)$, where $\delta(t)$ denotes the Kronecker delta function\cite{zheng2019radar};
$\mathbf{c}=\left[c[0],\cdots,c[N-1]\right]^{T}\in\mathbb{C}^{N\times 1}$ is known as the  {\it fast-time} code with $\left\|\mathbf{c}\right\|^2=N$, which is specially designed so that the waveform has good autocorrelation and spectrum property for sensing\cite{blunt2016overview}.
For example, for phase-coded waveforms, the typical fast-time codes include biphase Barker codes, minimum peak sidelobe (MPS) codes, maximal length sequence, Frank codes, and polyphase Barker codes~\cite{blunt2016overview}.
On the other hand, for the classic LFM pulse, where $p(t)=\exp\left(j\pi Bt^2/T_p\right)$, the fast-time code can be found as $c[n]=\exp(j\pi n^2/N),n=0,\cdots,N-1$.
Based on \eqref{x_k} and \eqref{p_t}, the transmitted waveform of HD pulse radar in \eqref{x_t} can be obtained accordingly.

\subsection{Pulsed Radar with Information Embedding}
The pulsed radar waveform discussed in Section~\ref{Conventional radar} is designed for radar sensing only.
To enable the additional information transmission yet without compromising the radar performance, an effective method is to embed the communication symbols into radar pulses, known as the radar-centric waveform for ISAC \cite{hassanien2016signaling}.
For example, one typical approach is to modulate radar pulses with phase-shift keying (PSK) symbols over the \emph{slow-time} scale \cite{hassanien2016signaling}, i.e., across PRIs\footnote{ Note that it is also possible to embed information symbols over fast-time code, i.e., over $c[0],\cdots,c[N-1]$, but at the cost of degrading the radar signal auto-correlation and spectrum properties.}.
In this case, the transmitted waveform for each PRI in \eqref{x_k} is modified as
\begin{equation}\label{x_psk}
x_k(t) = \left\{
\begin{split}
&\sqrt{P_r}\omega_k p (t),&0\le t\le T_p \\
&0,&T_p < t\le T
\end{split}, k=0,\cdots,K-1,
\right.
\end{equation}
where $\omega_k$ denotes the embedded $M$-ary PSK symbol that is selected from the signal constellation $\Omega=~\big\{\exp(j\frac{2\pi(m-1)}{M}),m=1,\cdots,M\big\}$.
It is noted that such a radar-centric waveform with PSK information embedding achieves information transmission without incurring any radar performance loss.
However, it can only support very low communication rate.
Specifically, for the given modulation order $M$, the communication spectrum efficiency is only
\begin{equation}
\mathcal{R}_{EB} = \frac{\log_2 M}{TB} = \frac{\rho\log_2 M}{T_pB} = \frac{\rho}{N}\log_2 M \ll \log_2 M \ \mathrm{bps/Hz},
\end{equation}
where the last inequality follows since $\rho\ll 1$ and $N\gg 1$ for typical pulsed radar waveforms \cite{richards2010principles}.

\subsection{Proposed FD-ISAC Waveform}
To overcome the above limitations on communication spectrum efficiency, while maintaining comparable or even achieving better sensing performance, we propose a novel FD-ISAC waveform design, as illustrated in Fig.~\ref{fd}.
With the proposed design, instead of remaining silence to wait for radar echoes, the FD-ISAC node transmits dedicated communication signals, so as to fully utilize the time duration $T$ for each PRI to enable high-rate communications.
Moreover, thanks to the FD operation that enables simultaneous transmission and reception, the echo reflected even from nearby target can be also captured by the receiver, as long as the SI is properly suppressed.
Note that SI can be mitigated by contemporary SIC techniques \cite{duarte2012experiment} and the fact that the FD-ISAC node perfectly knows the waveform transmitted by itself.
For the residual SI, proper power control over radar and communication signals can be applied to achieve desired performance.

With the proposed FD-ISAC scheme, the transmitted waveform $x_k(t)$ for the $k$-th PRI in \eqref{x_t} can be expressed as
\begin{equation}\label{fd_x}
x_k(t)= \left\{
\begin{split}
&\sqrt{P_r}\omega_k p(t),&&0\le t\le T_p \\
&\sqrt{P_c}g_k(t-T_p),&&T_p < t \le  T
\end{split}, k=0,\cdots,K-1,
\right.
\end{equation}
where $P_c$ is the transmit power of the dedicated communication signal that is in general different from the radar power $P_r$, and $g_k(t),0\le t \le T-T_p$, is the dedicated communication signal for the $k$-th PRI, which has the bandwidth $B$ and  can be expressed as
\begin{equation}
g_k(t) = \sqrt{T_c}\sum\limits_{j=0}^{J-1} s_k[j]\psi(t-jT_c),0\le t\le T-T_p,
\end{equation}
where $s_k[j]$ is the $j$-th dedicated communication symbol of the $k$-th PRI; $J=B(T-T_p)\gg 1$ denotes the number of dedicated communication symbols for each PRI.
As the communication symbols $s_k[j]$ are typically zero mean independent and identically distributed (i.i.d.) random variables (RVs) with normalized power, i.e., $\mathbb{E}[|s_k[j]|^2]=1$, it follows that $g_k(t)$ has the normalized power, i.e., $\mathbb{E}\left[\frac{1}{T-T_p}\int_0^{T-T_p}|g_k(t)|^2\mathrm{d}t\right]=~1$.

It is worth remarking that the proposed waveform in \eqref{fd_x} not only applies information embedding over the radar pulse as in \eqref{x_psk}, but also utilizes the otherwise wasted time $T-T_P\gg~T_p$ of each PRI for dedicated information transmission.
In practice, these two communication schemes could be used for different message types, e.g., information embedding for low-rate command and control (C\&C) messages and dedicated communication for high-capacity payload transmission (e.g., video, photo, etc.).
As such, the communication spectrum efficiency can be significantly improved.
For instance, if both information embedding and dedicated communication use $M$-order modulation, the communication spectrum efficiency of the proposed FD-ISAC waveform is
\begin{align}
\mathcal{R} &= \frac{\log_2M + J\log_2M}{TB} = \frac{\log_2M}{TB} + \frac{B(T-T_p)\log_2M}{TB} \notag\\
&= \frac{\rho}{N}\log_2M+\left(1-\rho\right)\log_2M \notag\\
&\approx \log_2M \gg \mathcal{R}_{EB}\ \ \mathrm{bps/Hz},
\end{align}
where the approximation in the last line follows from $\rho\ll 1$.
The above result shows that with the proposed FD-ISAC waveform, the communication spectrum efficiency not only significantly outperforms the existing information embedding scheme, but also approaches to $\log_2M$, which is the spectrum efficiency for pure communication system without radar signalling.
Such conclusion should come at no surprise considering that $\rho\ll 1$.
Furthermore, since $g_k(t)$ is the dedicated communication signal, its waveform can be more flexibly designed as compared to radar-centric information embedding.
Therefore, more efficient modulation can be used, say higher-order quadrature amplitude modulation (QAM), rendering higher communication spectrum efficiency.

While promising in principle, the performance of the proposed FD-ISAC scheme critically depends on the effectiveness of SIC techniques.
In the following, by taking into account the residual SI  for practical FD systems, the sensing and communication performances of the proposed FD-ISAC waveform are analyzed.

\section{Sensing Performance Analysis}\label{sensing_performance}
\subsection{Received Radar Echo}
First, we analyze the radar sensing performance for the proposed FD-ISAC scheme that takes into account the residual SI.
After appropriate clutter suppression \cite{richards2010principles}, the signal echo received at the FD-ISAC node over one CPI can be written as
\begin{equation}\label{received signal}
\begin{aligned}
y(t) &= \tilde{\alpha} x(t-\tau)e^{j2\pi f_d(t-\tau)} + \beta x(t) + n(t)
\\
& = \alpha x(t-\tau)e^{j2\pi f_dt} + \beta x(t) + n(t),0\le t\le KT,
\end{aligned}
\end{equation}
where $\alpha=\tilde{\alpha}e^{-j2\pi f_d\tau}$, with $\tilde{\alpha}$ denoting the complex channel coefficient of the reflected path by the radar target;
$\tau$ is the two-way propagation delay;
$f_d=\frac{2v_d}{\lambda}$ is the Doppler frequency shift caused by the target motion with the radial velocity $v_d$, where $\lambda$ is the carrier wavelength;
$\beta$ denotes the complex channel coefficient of the SI link, i.e., from the transmit antenna of the FD-ISAC node to its receive antenna;
$n(t)$ denotes additive white Gaussian noise (AWGN), which is assumed to be a circularly symmetric complex Gaussian (CSCG) random process, following $n(t)\sim\mathcal{CN}(0,N_0)$, with $N_0$ denoting the noise power spectrum density (PSD).
Note that in \eqref{received signal}, the propagation delay of the SI link is neglected since it is usually very small, or can be estimated offline and compensated appropriately.
It is worth remarking that for unambiguity range detection, the two-way propagation delay only in \eqref{received signal} needs to satisfy $0<\tau\le T-T_p$.
This is in a sharp contrast to the requirement of $T_p\le \tau\le T-T_p$ for conventional HD pulse radar, thanks to the FD capability of the proposed scheme.

By substituting \eqref{x_t} into \eqref{received signal}, we have
\begin{align}
y(t)&= \sum\limits_{k=0}^{K-1}\left[\alpha x_k(t-kT-\tau)e^{j2\pi f_d t} + \beta x_k(t-kT)\right] + n(t)
\notag\\
&= \sum\limits_{k=0}^{K-1}y_k(t-kT), 0 \le t\le KT,
\end{align}
with
\begin{align}\label{y_k}
y_k(t)=
&\alpha x_k(t-\tau)e^{j2\pi f_d t}e^{j2\pi f_d kT}+\beta x_k(t) \notag \\
&+n_k(t), 0\le t\le T,
\end{align}
where $n_k(t)$ is the AWGN with the same statistic as $n(t)$, but defined over one PRI only.
Note that within each PRI of duration $T$, the term $e^{j2\pi f_dt}$ in \eqref{y_k} can be omitted, since one single PRI duration $T$ is too short to cause notable phase variations with practical Doppler frequency, i.e., $f_dT\ll1$.
As a result, radar Doppler processing for velocity estimation must be performed over slow-time scale (say over one CPI) that contains many PRIs.
As a concrete example, consider an ISAC system at the carrier frequency $3.5$ GHz, which is also used for the 5G NR and has the great potential for ISAC \cite{TS38.101}.
A target with the radial velocity $120~\mathrm{km/h}$ would only cause Doppler frequency about $f_d\approx778~\mathrm{Hz}$.
With the PRI $T=10~\upmu s$, we have $f_dT=7.78\times 10^{-3}\ll 1$.
Therefore, within each PRI for $0\le t\le T$, we have $e^{j2\pi f_dt}\approx 1$.
Therefore, the resulting signal for each PRI $k$ in \eqref{y_k} can be written as
\begin{equation}\label{y_kt}
y_k(t) = \alpha x_k(t-\tau)e^{j2\pi f_dkT} + \beta x_k(t) + n_k(t), 0\le t\le T.
\end{equation}
Furthermore, based on $x_k(t)$ in \eqref{fd_x}, we have
\begin{equation}\label{x_k_tau}
\begin{aligned}
&x_k(t-\tau) = \left\{
\begin{aligned}
&x_{k-1}(t+T-\tau), && 0\le t\le \tau,\\
&x_k(t-\tau), && \tau< t \le \tau + T_p,\\
&x_k(t-\tau), &&\tau+T_p < t\le T
\end{aligned}
\right.
\\
&=\left\{
\begin{aligned}
&\sqrt{P_c}g_{k-1}(t+T-\tau-T_p),&&\ 0\le t\le \tau, \\
&\sqrt{P_r}\omega_k p(t-\tau),&& \tau < t \le \tau + T_p, \\
&\sqrt{P_c} g_k(t-\tau-T_p), && \tau + T_p< t \le T.
\end{aligned}
\right.
\end{aligned}
\end{equation}

\subsection{Signal Processing and Probability of Detection}
In the following, we analyze the radar sensing performance based on the standard radar processing procedures, as illustrated in Fig.\ref{flow}, which includes sampling, SIC, matched filtering (MF), Doppler processing, and target detection \cite{richards2010principles}.
\begin{figure*} \label{flow}
  \centering
  \includegraphics[width=0.9\textwidth]{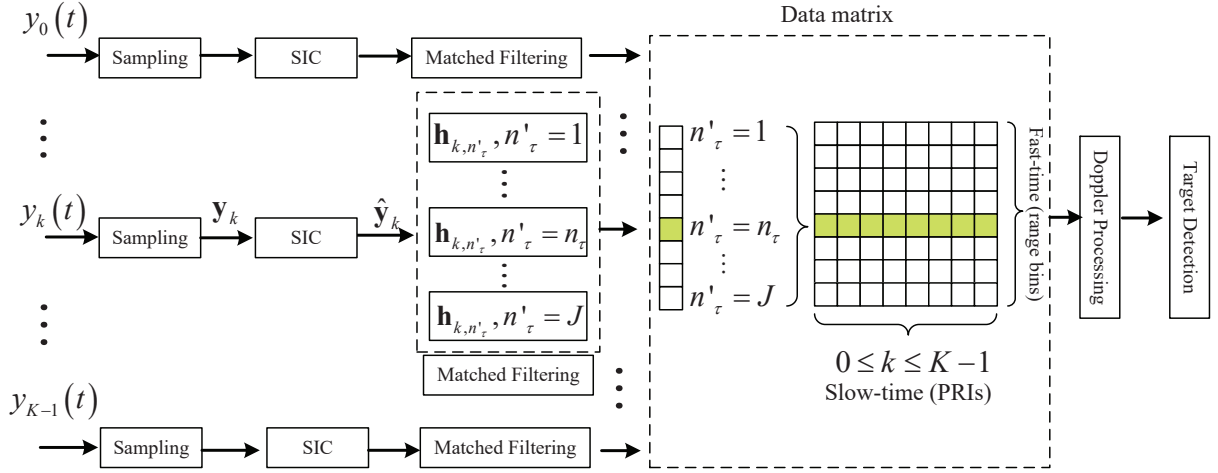}

  \caption{Radar signal processing with sampling, SIC, matched filtering, Doppler processing and target detection.}\label{flow}\vspace{-10pt}
\end{figure*}

\subsubsection{Sampling}
First, for each PRI $k$, the received waveform $y_k(t)$ in \eqref{y_kt} is projected onto the $(N+J)$ orthonormal basis functions $\left\{\psi(t-lT_c)\right\}_{l=0}^{N+J-1}$, so as to obtain an $(N+J)$ dimensional vector $\mathbf{y}_k=\left\{y_k[l]\right\}_{l=0}^{N+J-1}$, with
\begin{equation}\label{project}
y_k[l] = \left<y_k(t),\psi(t-lT_c)\right>,0\le l \le N+J-1,
\end{equation}
where $\left<a(t),b(t)\right>=\int_{-\infty}^{+\infty} a(t)b^*(t)\mathrm{d}t$ denotes the inner product of functions $a(t)$ and $b(t)$, and $\left(\cdot\right)^*$ denotes the complex conjugation.
By substituting \eqref{y_kt} into \eqref{project}, we have
\begin{equation}\label{y_k_l project}
y_k[l]
= \alpha e^{j2\pi f_dkT}x_{k,n_\tau}^{r}[l]+ \beta x_k^d[l] + n_k[l],
\end{equation}
where $x_{k,n_\tau}^r[l]\triangleq\left<x_k(t-\tau),\psi(t-lT_c)\right>$ corresponds to the projected version of $x_k(t-\tau)$, whose expression is to be derived later in \eqref{project x_k};
$x_k^{d}[l] \triangleq\left<x_k(t),\psi(t-lT_c)\right>$, which is given by
\begin{equation}\label{sid}
x_k^{d}[l]=\left\{
\begin{aligned}
&\sqrt{P_rT_c} \omega_k c[l],&& 0\le l\le N-1, \\
&\sqrt{P_cT_c}s_k[l-N],&&\ N\le l \le N+J-1.
\end{aligned}
\right.
\end{equation}
\begin{IEEEproof}
Please refer to  Appendix \ref{appendix a}.
\end{IEEEproof}
Furthermore, the projected noise in \eqref{y_k_l project} is $n_k[l]\triangleq\left<n_k(t),\psi(t-lT_c)\right>$, which is given by
\begin{equation}\label{n_k}
n_k[l]=\int_{0}^{T}n_k(t)\psi^*(t-lT_c)\mathrm{d}t, 0\le l \le N+J-1.
\end{equation}
It can be shown that $n_k[l]\overset{i.i.d.}{\sim}\mathcal{CN}(0,N_0)$, for $k=0,\cdots,K-1$ and $l=~0,\cdots,N+J-1$.

To derive $x_{k,n_\tau}^r[l]\triangleq\left<x_k(t-\tau),\psi(t-lT_c)\right>$ in \eqref{y_k_l project}, we first note that for the signal with bandwidth $B$, the time resolution for sensing is $T_c=1/B$.
Therefore, let the delay $\tau\approx~ n_\tau T_c$, where $n_\tau=~\mathrm{round}(\tau B)$ is a positive integer, with $\mathrm{round}(\cdot)$ denoting round to the nearest integer.
As $0<\tau\le T-T_p$, we have $1\le n_\tau\le J,$
which correspond to the detectable {\it range bins}.
According \eqref{x_k_tau}, for a given delay $\tau$ or $n_\tau$, we can obtain that
\begin{equation}\label{project x_k}
\begin{array}{l}
x_{k,n_\tau}^{r}[l]\triangleq\left<x_k(t-\tau),\psi(t-lT_c)\right> \\
=\left\{
\begin{split}
&\sqrt{P_cT_c} s_{k-1}[J-n_\tau+l],&& 0\le l \le n_\tau-1, \\
&\sqrt{P_rT_c} w_k c[l-n_\tau],&& n_\tau \le l \le N+n_\tau -1, \\
&\sqrt{P_cT_c} s_k[l-N-n_\tau],&& N+n_\tau \le l \le N + J - 1.
\end{split}
\right.
\end{array}
\end{equation}
\begin{IEEEproof}
Please refer to  Appendix \ref{appendix b}.
\end{IEEEproof}
Therefore, \eqref{y_k_l project} can be written compactly in vector form as
\begin{equation}\label{y_k_final}
\begin{split}
\mathbf{y}_k =\alpha e^{j2\pi f_d kT}\mathbf{x}_{k,n_\tau}^r + \beta\mathbf{x}_k^d + \mathbf{n}_k,
\end{split}
\end{equation}
where $\mathbf{y}_k\in\mathbb{C}^{(N+J)\times 1}$; $\mathbf{n}_k$ is the CSCG random vector with $\mathbf{n}_k\sim~\mathcal{CN}(0,N_0\mathbf{I}_{N+J})$; $\mathbf{x}_{k,n_\tau}^r=~\mathrm{diag}\left(\sqrt{T_c\mathbf{p}^{r}_{n_\tau}}\right)\mathbf{r}_{k,n_\tau}$ and $\mathbf{x}_k^d=\mathrm{diag}\left(\sqrt{T_c\mathbf{p}^{d}}\right)\mathbf{d}_{k}$ denote the radar echo and SI vectors, respectively, with
\begin{equation}\label{component}
\begin{aligned}
&\mathbf{p}_{n_\tau}^{r}=[P_c\mathbf{1}_{n_\tau},P_r\mathbf{1}_N,P_c\mathbf{1}_{J-n_\tau}]\in\mathbb{R}^{1\times (N+J)}
\\
&
\begin{aligned}
\mathbf{r}_{k,n_\tau}=[s_{k-1}[J-n_\tau],\cdots,s_{k-1}[J-1],\omega_k\mathbf{c},s_k[0],
\cdots,&\\
\cdots,s_k[J-n_\tau-1]]^T\in\mathbb{C}^{(N+J)\times 1}&
\end{aligned}
\\
&\mathbf{p}^{d}=[P_r\mathbf{1}_{N},P_c\mathbf{1}_J]\in\mathbb{R}^{1\times (N+J)}
\\
&\mathbf{d}_k=\left[\omega_k\mathbf{c},s_k[0],s_k[1],\cdots,s_k[J-1]\right]^{T}\in\mathbb{C}^{(N+J)\times 1}
\end{aligned}
\end{equation}
with $\mathbf{1}_N$ denoting the $1\times N$ all-one vector.
\subsubsection{SIC}
Since the transmitted waveform is known at the radar receiver due to the monostatic architecture, the resulting SI can be in principle cancelled by subtracting $\beta \mathbf{x}_k^d$ from $\mathbf{y}_k$ in \eqref{y_k_final}.
However, in practice, since SI is usually orders-of-magnitude stronger than the useful echo signal, the receiver frontend, such as the low noise amplifier and analog-to-digital converter (ADC), may become saturated before effective analog/digital SIC techniques can be applied, which introduces nonlinearity distortion.
Therefore, the residual SI still exists after several stages of SIC processes, e.g., antenna separation, analog cancellation, and digital cancellation \cite{duarte2012experiment}, which is usually modelled as the worst case Gaussian noise, whose power is proportional to the input SI power before cancellation \cite{duarte2012experiment}.
Therefore, the resulting signal of \eqref{y_k_final} after SIC can be written as
\begin{equation}\label{sic}
\begin{split}
&\hat{\mathbf{y}}_k = \alpha e^{j2\pi f_d kT}\mathbf{x}_{k,n_\tau}^r
+\sqrt{\epsilon}\beta\mathrm{diag}(\sqrt{T_c\mathbf{p}^{d}})\mathbf{z}_{k} + \mathbf{n}_k,
\end{split}
\end{equation}
where the second term models the residual SI with $\epsilon\ll 1$ denoting the extent of SIC, and $\mathbf{z}_k$ is a CSCG random vector with $\mathbf{z}_k\sim\mathcal{CN}(0,\mathbf{I}_{N+J})$.

\subsubsection{MF}
It is well-known that MF is optimal for maximizing the output SNR for radar detection \cite{richards2010principles}.
In practice, MF can be implemented in analog or digital domain via various techniques, such as convolution or correlation.
With the resulting signal \eqref{sic}, the effect of MF can be studied by applying $J$ parallel filters $\mathbf{h}_{k,n_\tau'}$, each corresponding to one range bin $n_\tau'$ where the target might lie in.
Specifically, the MF vector can be written as
\begin{equation}\label{mfer}
\mathbf{h}_{k,n_\tau'}=\frac{\mathbf{x}_{k,n_\tau'}^r}{\left\|\mathbf{x}_{k,n_\tau'}^r\right\|}, 1\le n_\tau'\le J.
\end{equation}
Of particular interest is the probability of detection for the particular range-Doppler bin where the target actually lies.
For other range-Doppler bins, similar analysis can be applied.
For the particular range bin that matches with the true radar delay, i.e., $n_\tau'=n_\tau$, the output of the MF is
\begin{equation}\label{mf}
\begin{aligned}
&\mathrm{y_k} = \mathbf{h}_{k,n_\tau}^\mathrm{H}\hat{\mathbf{y}}_{k} \\
&=\alpha e^{j2\pi f_d kT}\left\|\mathbf{x}_{k,n_\tau}^r\right\|+\frac{\mathbf{x}_{k,n_\tau}^{r\ \mathrm{H}}\left[\sqrt{\epsilon}\beta\mathrm{diag}(\sqrt{T_c\mathbf{p}^{d}})\mathbf{z}_{k} \right]}{\left\|\mathbf{x}_{k,n_\tau}^r\right\|}+\mathrm{n}_k,
\end{aligned}
\end{equation}
where $\mathrm{n}_k=\mathbf{h}_{k,n_\tau}^\mathrm{H}\mathbf{n}_k\sim\mathcal{CN}(0,N_0)$.
Furthermore, we have
\begin{equation}\label{x_k_norm}
\begin{aligned}
&\left\|\mathbf{x}_{k,n_\tau}^r\right\| = \left\|\mathrm{diag}\left(\sqrt{T_c\mathbf{p}^{r}_{n_\tau}}\right)\mathbf{r}_{k,n_\tau}\right\|
\\
&=\sqrt{P_cT_c\left(\sum\limits_{j=J-n_\tau}^{J-1}|s_{k-1}[j]|^2+\sum\limits_{j=0}^{J-n_\tau-1}|s_k[j]|^2\right)+P_rT_cN}.
\end{aligned}
\end{equation}
Note that \eqref{x_k_norm} involves the summation of $J$ i.i.d. random communication symbols with normalized power.
When $J$ is large, based on the law of large numbers, we have
\begin{equation}\label{lln}
\left\|\mathbf{x}_{k,n_\tau}^r\right\|\rightarrow \sqrt{T_c\left(P_cJ+P_rN\right)}.
\end{equation}
By substituting \eqref{lln} into \eqref{mf}, we have
\begin{equation}\label{mf2}
\mathrm{y}_k = \alpha e^{j2\pi f_dkT}\sqrt{T_c\left(P_cJ+P_rN\right)} + \mathrm{z}_k + \mathrm{n}_k,
\end{equation}
where
\begin{equation}\label{z_f}
\mathrm{z}_k = \frac{\sqrt{\epsilon}\beta T_c\mathbf{r}_{k,n_\tau}^\mathrm{H}\mathrm{diag}(\sqrt{\mathbf{p}_{n_\tau}^r\circ\mathbf{p}^d})\mathbf{z}_k}{\sqrt{T_c\left(P_cJ+P_rN\right)}},
\end{equation}
with $\circ$ denoting the Hadamard product.
According to \eqref{component}, we can obtain that
\begin{equation}\label{range_case}
\begin{aligned}
&\mathbf{p}_{n_\tau}^r\circ\mathbf{p}^d=
\\&\left\{
\begin{aligned}
&\left[P_cP_r\mathbf{1}_{n_\tau},P_r^2\mathbf{1}_{N-n_\tau},P_rP_c\mathbf{1}_{n_\tau},P_c^2\mathbf{1}_{J-n_\tau}\right], 1\le n_\tau\le N, \\
&\left[P_cP_r\mathbf{1}_{N},P_c^2\mathbf{1}_{n_\tau-N},P_rP_c\mathbf{1}_{N},P_c^2\mathbf{1}_{J-n_\tau}\right], \ N<n_\tau\le J.
\end{aligned}
\right.
\end{aligned}
\end{equation}
Note that the residual SI power in \eqref{range_case} share similar pattern for the above two cases, except the difference for the second block, where we have $P_r^2$ versus $P_c^2$.
This is expected due to the different target range considered.
For nearby target such that $1\le n_\tau\le N$, the first echo arrives while the FD-ISAC node is still transmitting the radar pulse with power $P_r$.
On the other hand, for $N<n_\tau\le J$, the first target echo arrives while the FD-ISAC node is transmitting the dedicated communication signal with power $P_c$.
It follows from \eqref{z_f} and \eqref{range_case} that the residual SI power after MF is
\begin{equation}\label{z_k}
\begin{aligned}
&\mathbb{E}\left[|\mathrm{z}_k|^2\right] = \frac{\mathbb{E}\left[\left|\sqrt{\epsilon}\beta T_c\mathbf{r}_{k,n_\tau}^\mathrm{H}\mathrm{diag}\left(\sqrt{\mathbf{p}_{n_\tau}^r\circ\mathbf{p}^d}\right)\mathbf{z}_{k}\right|^2\right]}{T_c\left(P_cJ+P_rN\right)}
\\
&=\left\{
\begin{aligned}
&\frac{\epsilon|\beta|^2T_c\left(P_r^2N+P_c^2J-n_\tau\left(P_r-P_c\right)^2\right)}{P_cJ+P_rN}, 1\le n_\tau\le N,\\
&\frac{\epsilon|\beta|^2T_c\left(P_c^2\left(J-N\right)+2P_cP_rN\right)}{P_cJ+P_rN}, \quad\quad\quad N<n_\tau\le J.
\end{aligned}
\right.
\end{aligned}
\end{equation}
Therefore, it follows from \eqref{mf2} that the average signal-to-interference-plus-noise ratio (SINR) of the radar echoes for one PRI can be obtained as
\begin{equation}\label{sinr_1}
\begin{array}{l}
\mathrm{SINR}_1 = \frac{\left|\alpha e^{j2\pi f_dkT}\sqrt{T_c\left(P_cJ+P_rN\right)}\right|^2 }{\mathbb{E}\left[\left|\mathrm{z}_k+\mathrm{n}_k\right|^2\right]}
=
\frac{\left|\alpha\right|^2T_c\left(P_cJ+P_rN\right)}{\mathbb{E}\left[\left|\mathrm{z}_k\right|^2\right]+N_0}.
\end{array}
\end{equation}
By substituting \eqref{z_k} into \eqref{sinr_1}, we obtain the SINR as a function of the transmit power $P_r$ and $P_c$ as
\begin{equation}\label{sinr_1_final}
\begin{aligned}
&\mathrm{SINR}_1(P_r,P_c)\\
&=\left\{
\begin{aligned}
&\frac{|\alpha|^2\left(P_cJ+P_rN\right)}{\frac{\epsilon|\beta|^2\left(P_r^2N+P_c^2J-n_\tau(P_r-P_C)^2\right)}{P_cJ+P_rN}+N_0B}, 1\le n_\tau\le N,
\\ &\frac{|\alpha|^2\left(P_cJ+P_rN\right)}{\frac{\epsilon|\beta|^2\left(P_c^2(J-N)+2P_rP_cN\right)}{P_cJ+P_rN}+N_0B}, \quad\quad N<n_\tau\le J.
\end{aligned}
\right.
\end{aligned}
\end{equation}
Note that \eqref{sinr_1_final} is an unified expression that takes into account the residual SI, which includes the pulsed radar and CW radar as special cases, as elaborated in the following:

When $P_c=0$, the resulting SINR in \eqref{sinr_1_final} reduces to
\begin{equation}\label{sinr_case1}
\mathrm{SINR}_1(P_r,0)=\left\{
\begin{split}
&\frac{\left|\alpha\right|^2 P_rN}{\frac{\epsilon\left|\beta\right|^2P_r(N-n_\tau)}{N}+N_0B},&& 1\le n_\tau\le N, \\
&\frac{\left|\alpha\right|^2 P_rN}{N_0B},&& N<n_\tau\le J,
\end{split}
\right.
\end{equation}
where the proposed FD-ISAC waveform reduces to the conventional pulsed radar waveform, but has the capability for nearby target sensing due to the FD operation.
Here, we comment that for $1\le n_\tau\le N$, the output SINR is affected by the residual SI power; while for $N<n_\tau\le J$, the target echo arrives after the radar pulse transmission is completed, so that no SI incurs.
With perfect SIC, i.e., $\epsilon=0$, the SINR expressions of both cases in \eqref{sinr_case1} are identical.

When $P_c=P_r=\bar{P}$, where $\bar{P}$ denotes the average transmit power, the resulting SINR in \eqref{sinr_1_final} reduces to
\begin{equation}\label{CW-SINR}
\mathrm{SINR}_1(\bar{P},\bar{P})=\frac{\left|\alpha\right|^2 \bar{P}(J+N)}{\epsilon\left|\beta\right|^2\bar{P}+N_0B}, 1\le n_\tau\le J.
\end{equation}
Note that \eqref{CW-SINR} is applicable for any CW with constant envelope that takes into account the residual SI, including the classic FMCW waveform.
While conventional FMCW sensing usually ignores the residual SI by assuming nearby target sensing with small $\bar{P}$, as the transmit power increases for long range target detection, the effect of residual SI should be taken into account by using \eqref{CW-SINR}.
When the SIC capability is poor, such CW schemes suffers from serious residual SI power.
By contrast, our proposed FD-ISAC scheme can flexibly control the transmit power to mitigate the residual SI effect by reducing $P_c$ while increasing $P_r$, to achieve desired SINR for sensing.

When $P_r=0$, the resulting SINR in \eqref{sinr_1_final} reduces to
\begin{equation}\label{P_r_0}
\mathrm{SINR}_1(0,P_c)=\left\{
\begin{aligned}
&\frac{\left|\alpha\right|^2 P_cJ}{\frac{\epsilon\left|\beta\right|^2P_c(J-n_\tau)}{J}+N_0B},&& 1\le n_\tau\le N, \\
&\frac{\left|\alpha\right|^2 P_cJ}{\frac{\epsilon\left|\beta\right|^2P_c(J-N)}{J}+N_0B},&& \ N<n_\tau\le J.
\end{aligned}
\right.
\end{equation}
In this case, no dedicated radar waveform is transmitted, and sensing is achieved based on the communication signal.

\subsubsection{Doppler processing}
As illustrated in Fig.~\ref{flow}, after MF, for all the $J$ range bins over the $K$ PRIs in each CPI, we obtain a data matrix of dimension $J\times K$.
Based on the standard radar signal processing \cite{richards2010principles}, for each range bin, i.e., each row of the data matrix in Fig.~\ref{flow}, Doppler processing is applied over the slow-time data for estimating the Doppler frequency $f_d$, and hence the target radial velocity $v_d$.

In essence, Doppler processing is to compute the discrete Fourier transform (DFT) of the slow-time data for each range bin, which is periodic in frequency with principal period ranging from $-\frac{1}{2T}$ to $\frac{1}{2T}$, so that the Doppler shift within $\left|f_d\right|\le \frac{1}{2T}$ can be unambiguously detected \cite{richards2010principles}.
Let the ground truth Doppler frequency be represented as $f_d\approx\frac{q}{KT}$ for some $q$, where $\frac{1}{KT}$ is the Doppler resolution \cite{richards2010principles}, and $q\in\left\{-\frac{K}{2},\cdots,-1,1,\cdots,\frac{K}{2}\right\}$ corresponds to the Doppler bin.
Note that the negative or positive $f_d$ corresponds to target moving away from or towards the radar receiver, respectively.
For the particular range bin $n_\tau$ where the target lies, the DFT Doppler processing is performed over the MF output $\mathrm{y}_k, k=0,\cdots,K-1$ in \eqref{mf2}, which yields
\begin{align}\label{dft}
&Y[q'] = \frac{1}{\sqrt{K}}\sum\limits_{k=0}^{K-1}\mathrm{y}_k e^{-j2\pi\frac{q'}{K}k},  q'= -\frac{K}{2},\cdots,-1,1,\cdots,\frac{K}{2} \notag\\
&= \frac{1}{\sqrt{K}}\sum\limits_{k=0}^{K-1}\left(\alpha e^{j2\pi f_dkT}\sqrt{\mathcal{E}} + \mathrm{z}_k + \mathrm{n}_k\right)e^{-j2\pi\frac{q'}{K}k}
\notag\\
&\approx\frac{1}{\sqrt{K}}\sum\limits_{k=0}^{K-1}\left(\alpha e^{j2\pi \frac{q}{K}k}\sqrt{\mathcal{E}} + \mathrm{z}_k + \mathrm{n}_k\right)
e^{-j2\pi\frac{q'}{K}k}.
\end{align}
where $\mathcal{E}\triangleq T_c\left(P_cJ+P_rN\right)$.
For the particular Doppler bin with $q'=q$, the peak of \eqref{dft} can be obtained as
\begin{equation}\label{dft2}
Y[q] = \alpha\sqrt{K\mathcal{E}}+\varphi = \alpha\sqrt{KT_c\left(P_cJ+P_rN\right)}+\varphi,
\end{equation}
where $\varphi=\frac{1}{\sqrt{K}}\sum\limits_{k=0}^{K-1}\left(\mathrm{z}_k+\mathrm{n}_k\right)e^{-j2\pi\frac{q}{K}k}$ is the resulting noise after the DFT operation, which is a CSCG RV with zero mean and variance  $\sigma_\varphi^2=\mathbb{E}[\left|\varphi\right|^2]=\mathbb{E}[\left|\mathrm{z}_k\right|^2]+N_0$.
Base on \eqref{dft2}, after Doppler processing over slow-time data, the average SINR of radar echoes at the peak corresponding to the particular range-Doppler bin where the target lies is
\begin{equation}\label{SINR_K}
\mathrm{SINR}_K = \frac{\left|\alpha\right|^2KT_c\left(P_cJ+P_rN\right)}{\mathbb{E}\left[\left|\mathrm{z}_k\right|^2\right]+N_0}
=K\cdot\mathrm{SINR}_1,
\end{equation}
where the last equality follows from \eqref{sinr_1}.
It is observed that with Doppler processing over the $K$ PRIs of each CPI, the peak SINR is increased by $K$ times relative to the SINR for one PRI.
This is known as the {\it coherent integration gain} of the slow-time data \cite{blunt2016overview,richards2010principles}.
\subsubsection{Target Detection}\label{detection}
After the procedures discussed above, target detection is performed.
Note that in many practical applications, a detection decision needs to be made for each range-Doppler bin.
Of particular interest is the detection probability of the range-Doppler bin $(n_\tau,q)$ where the target actually lies.
Based on \eqref{dft2}, depending on whether the target exists or not, we can obtain the signal output as
\begin{equation}\label{hypothesses}
\mathcal{H}_1:Y = \alpha\sqrt{KT_c(P_cJ+P_rN)}+\varphi,\quad\mathcal{H}_0:Y=\varphi,
\end{equation}
which forms a problem of binary hypothesis testing.
In practice, the {\it linear detector} is usually used \cite{richards2010principles}, i.e.,
\begin{equation}
z=|Y|\overset{\mathcal{H}_1}{\underset{\mathcal{H}_0}{\gtrless}}\mathcal{T},
\end{equation}
where $\mathcal{T}$ denotes the predetermined threshold.
According to \eqref{hypothesses}, since $\varphi$ is a CSCG RV, $z$ under hypothesis $\mathcal{H}_0$ satisfies Rayleigh distribution as
\begin{equation}
p_z(z\mid\mathcal{H}_0) = \left\{
\begin{split}
&\frac{2z}{\sigma_{\varphi}^2}\exp\left(-\frac{z^2}{\sigma_{\varphi}^2}\right),&& z\ge 0 \\
&0,&& z<0,
\end{split}
\right.
\end{equation}
and the probability of false alarm can be derived as
\begin{equation}\label{P_fa}
P_{FA} = \int_{\mathcal{T}}^{\infty}p_z(z\mid\mathcal{H}_0)\mathrm{d}z = \exp\left(-\frac{\mathcal{T}}{\sigma_{\varphi}^2}\right).
\end{equation}
Therefore, by fixing the probability of false alarm to a constant, we can derive the threshold $\mathcal{T}$ from \eqref{P_fa} as $\mathcal{T} =~ \sigma_{\varphi}\sqrt{-\ln P_{FA}}$.
Note that different from conventional radar where the noise power is usually modelled as a constant for all range bins, $\sigma_\varphi$ involves not just the background noise, but also the residual SI, whose power is dependent on the delay $n_\tau$ as evident from \eqref{z_k} and \eqref{SINR_K}.
Fortunately, as the detection decision is made for each range-Doppler bin,
threshold $\mathcal{T}$ can be adjusted accordingly based on $\sigma_\varphi$  to maintain a constant $P_{FA}$ \cite{richards2010principles}.

On the other hand, when the target exists, $z$ under hypothesis $\mathcal{H}_1$ satisfies Rician distribution as
\begin{equation}\label{p_h1}
p_{z}(z\mid \mathcal{H}_1) = \left\{
\begin{aligned}
&\frac{2z}{\sigma_{\varphi}^2}\exp\left[-\frac{(z^2+m^2)}{\sigma_{\varphi}^2}\right]I_0\left(\frac{2m^2z}{\sigma_\varphi}\right),&& z\ge0,\\
&0,&& z<0,
\end{aligned}
\right.
\end{equation}
where $m\triangleq|\alpha|\sqrt{KT_c(P_cJ+P_rN)}$, and the probability of detection can be obtained as
\begin{equation}\label{pd}
P_D = \int_{\mathcal{T}}^{\infty} p_z(z\mid\mathcal{H}_1)\mathrm{d}z.
\end{equation}
By substituting \eqref{p_h1} into \eqref{pd}, for any desired false alarm rate $P_{FA}$, the probability of detection can be derived as \cite[Chap. 15 Eq. (45)]{richards2010principles}
\begin{equation}\label{p_d}
\begin{aligned}
&P_D(P_r,P_c,n_\tau) = Q_1\left(\sqrt{\frac{2m^2}{\sigma_\varphi^2}},\sqrt{\frac{2\mathcal{T}^2}{\sigma_\varphi^2}}\right) \\
&=Q_1\left(\sqrt{\frac{2|\alpha\sqrt{KT_c(P_cJ+P_rN)}|^2}{\mathbb{E}[|\varphi|^2]}},\sqrt{\frac{2\sigma_\varphi^2(-\ln P_{FA})}{\sigma_\varphi^2}}\right) \\
&=Q_1\left(\sqrt{2\cdot\mathrm{SINR}_K},\sqrt{-2\ln P_{FA}}\right)\\
&=Q_1\left(\sqrt{2K\cdot\mathrm{SINR}_1(P_r,P_c,n_\tau)},\sqrt{-2\ln P_{FA}}\right),
\end{aligned}
\end{equation}
where $Q_1(a,b)$ denotes the first-order Marcum $Q$-function with parameters $a$ and $b$.
Note that the probability of detection monotonically increases with the output SINR, while independent of the transmitted waveform.
Therefore, \eqref{p_d} is a unified expression of the probability of detection for monostatic sensing that takes into account the residual SI, which is suitable for either pulsed or CW radars, including pulsed LFM, FMCW or communication-centric radars.

\subsection{Ambiguity Function Analysis}
In addition to probability of detection, AF is another important performance metric for sensing, which characterizes the sensing resolution capability for range and Doppler estimation.
For the $k$-th PRI, the normalized AF for the transmit waveform $x_k(t)$ can be expressed as
\begin{equation}\label{AF}
\chi_n(\tau,f_d) = \frac{\chi_k\left(\tau,f_d\right)}{\chi_k\left(0,0\right)},
\end{equation}
with
\begin{equation}\label{AF-0}
\chi_k\left(\tau,f_d\right)\triangleq\left|\int_{0}^{T}x_k(t)x_k^*(t-\tau)e^{-j2\pi f_dt}\mathrm{d}t\right|,
\end{equation}
where as discussed for \eqref{y_k} and \eqref{y_kt}, since $f_dT\ll 1$ for practical Doppler frequency $f_d$ and PRI $T$, we have $e^{-j2\pi f_dt}\approx 1$, for all $0\leq t\leq T$.
Therefore, \eqref{AF-0} can be approximated as
\begin{equation}\label{ACF}
\chi_k\left(\tau,f_d\right)\approx\chi_k\left(\tau,0\right) = \left|\int_{0}^{T}x_k(t)x_k^*(t-\tau)\mathrm{d}t\right|,
\end{equation}
where $\chi_k\left(\tau,0\right)$ is the ACF of $x_k(t)$.
Therefore, the range estimation performance is mainly determined by the autocorrelation property for signal $x_k(t)$ within each PRI, while that for Doppler estimation is mainly related to the waveform across the $K$ PRIs.
Therefore, we mainly focus on the ACF for range estimation performance.
Since ACF is symmetric with respect to the origin, we only consider $0\le \tau\le T-T_p$.
By substituting \eqref{fd_x} and \eqref{x_k_tau} into \eqref{ACF}, we can obtain that
\begin{equation}
\chi_k(\tau,0) =
\left\{
\begin{aligned}
&P_r\chi_p(\tau,0)+\Phi(P_r,P_c,\tau),&&0\le\tau\le T_p,\\
&\Psi(P_r,P_c,\tau),&&T_p<\tau\le T-T_p,
\end{aligned}
\right.
\end{equation}
where $\chi_p(\tau,0)\triangleq\int_{\tau}^{T_p}p(t)p^*(t-\tau)\mathrm{d}t$ denotes the ACF of the radar pulse, which typically has good and constant autocorrelation property for range sensing, and when $\tau=0$, we have $\chi_p(0,0)=\int_0^{T_p}\left|p(t)\right|^2\mathrm{d}t=T_P$;
Furthermore, $\Phi(P_r,P_c,\tau)$ and $\Psi(P_r,P_c,\tau)$ denote
the summation of the cross-correlation function (CCF) between radar pulse and dedicated communication signals and the ACF of the dedicated communication signals, which are given by
\begin{align}
&\Phi(P_r,P_c,\tau)\triangleq\sqrt{P_rP_c}\int_{0}^{\tau}\omega_kp(t)g_{k-1}^*(t+T-\tau-T_p)\mathrm{d}t
\notag\\
&\qquad+\sqrt{P_rP_c}\int_{T_p}^{\tau+T_p}g_{k}(t-T_p)\omega_k^*p^*(t-\tau)\mathrm{d}t\notag\\
&\qquad+P_c \int_{\tau+T_p}^{T}g_{k}(t-T_p)g_k^*(t-\tau-T_p)\mathrm{d}t\\
&\Psi(P_r,P_c,\tau)\triangleq\sqrt{P_rP_c}\int_{0}^{T_p}\omega_kp(t)g_{k-1}^*(t+T-\tau-T_p)\mathrm{d}t\notag\\
&\qquad+P_c\int_{T_p}^{\tau}g_k(t-T_p)g_{k-1}^*(t+T-T_p-\tau)\mathrm{d}t\notag\\
&\qquad+\sqrt{P_rP_c}\int_{\tau}^{T_p+\tau}g_k(t-T_p)\omega_k^*p^*(t-\tau)\mathrm{d}t\notag\\
&\qquad+P_c\int_{\tau+T_p}^{T}g_k(t-T_p)g_{k-1}^*(t-\tau-T_p)\mathrm{d}t.
\end{align}
When $\tau=0$, we can obtain that
\begin{align}
&\Phi(P_r,P_c,0)=P_c\int_{T_p}^{T}\left|g_k(t-T_p)\right|^2\mathrm{d}t=P_c(T-T_p),\notag\\
&\Psi(P_r,P_c,0)=0.
\end{align}
Therefore, we have
$
\chi_k(0,0)=P_rT_p+P_c(T-T_p),
$
and the normalized ACF can be derived as
\begin{equation}\label{ACF-N}
\begin{aligned}
&\chi_n(\tau,0)={\chi_k(\tau,0)}/{\chi_k(0,0)}\\
&=\left\{
\begin{aligned}
&\frac{P_r\chi_p(\tau,0)+\Phi(P_r,P_c,\tau)}{P_rT_p+P_c(T-T_p)},&&0\le\tau\le T_p,\\
&\frac{\Psi(P_r,P_c,\tau)}{P_rT_p+P_c(T-T_p)},&&T_p\le\tau\le T-T_p.
\end{aligned}
\right.
\end{aligned}
\end{equation}
Note that since $\Phi(P_r,P_c,\tau)$ and $\Psi(P_r,P_c,\tau)$ are random functions influenced by random communication symbols $\omega_k$ and $g_k(t)$, different from the conventional radar waveforms, in ISAC systems, the ACF is a random function.
For communication-centric radars, the random communication waveforms may lead to high instantaneous peak-to-sidelobe level (PSL) for the ACF, rendering poor range sensing performance.
By contrast, for our proposed FD-ISAC scheme, if better sensing capability is desired in terms of the ACF, the transmit power can be flexibly adjusted by reducing the dedicated communication power $P_c$, while concentrating more power on radar pulse.
As an extreme example, when $P_c=0$, we can obtain that $\Phi(P_r,0,\tau)=\Psi(P_r,0,\tau)=0$, and \eqref{ACF-N} reduces to the normalized ACF for the classic pulsed radars with superior autocorrelation property for range sensing.

\section{Communication Performance Analysis}\label{communication_performance}
In this section, we analyze the communication performance of the proposed FD-ISAC scheme.
The received signal at the communication receiver can be written as
\begin{equation}\label{y_c}
y_c(t) = hx(t) + n(t), 0\le t \le KT,
\end{equation}
where $h$ denotes the complex channel coefficient between the FD-ISAC node and the communication receiver.
By substituting \eqref{x_t} into \eqref{y_c}, we can obtain that
\begin{equation}
\begin{aligned}
y_c(t) &= \sum\limits_{k=0}^{K-1} hx_k(t-kT) + n(t) \\
&=\sum\limits_{k=0}^{K-1}y_{c,k}(t-kT), 0\le t\le KT,
\end{aligned}
\end{equation}
where $y_{c,k}(t) = hx_k(t) + n_k(t), 0\le t \le T$.
Based on the definition of $x_k(t)$ in \eqref{fd_x}, the received signal at the communication receiver for each PRI $k$ is
\begin{equation}\label{y_c_r}
y_{c,k}(t) = \left\{
\begin{split}
&h\sqrt{P_r}\omega_k p(t) + n_k(t),&&0\le t\le T_p, \\
&h\sqrt{P_c}g_k(t-T_p)+n_k(t),&&T_p < t \le  T,
\end{split}
\right.
\end{equation}
which consists of the radar pulse embedded with communication symbols $\omega_k$ and the dedicated communication signal.
After  ADC, the discrete-time equivalent of \eqref{y_c_r} is given by
\begin{equation}\label{y_c_k}
y_{c,k}[l] = \left\{
\begin{split}
&h\sqrt{P_rT_c}\omega_kc[l]+n_k[l],&& 0\le l\le N-1, \\
&h\sqrt{P_cT_c}s_k[l-N]+n_k[l],&& N\le l\le N+J-1,
\end{split}
\right.
\end{equation}
where $k=0,\cdots,K-1$.

First, we study the performance of the embedded PSK communications, i.e., $y_{c,k}[l], 0\le l\le N-1$.
With the radar fast-time code $c[l], 0\le l \le N-1$, known at the communication receiver, the embedded PSK communication symbols can be demodulated by applying MF as
\begin{equation}\label{psk_decode}
\begin{split}
\mathrm{y}_{c,k}&= \frac{1}{\sqrt{N}}\sum_{l=0}^{N-1}c^*[l]y_{c,k}[l]
\\
&
=\frac{1}{\sqrt{N}}\sum_{l=0}^{N-1}c^*[l]\left(h\sqrt{P_rT_c}\omega_kc[l]+n_k[l]\right)\\
&=h\sqrt{P_rT_cN}\omega_k + \mathrm{n}_{k,c},
\end{split}
\end{equation}
where $\mathrm{n}_{k,c}=\frac{1}{\sqrt{N}}\sum_{l=0}^{N-1}c^*[l]n_k[l]$ with $\mathrm{n}_{k,c}\sim\mathcal{CN}(0,N_0)$.
The resulting output SNR is $\frac{|h|^2P_rNT_c}{N_0}=\frac{|h|^2P_rN}{N_0B}$, where the factor $N$ corresponds to the processing gain due to the radar fast-time code.
Therefore, for  $M$-ary PSK signalling, the probability of symbol error is \cite[Chap. 6, Eq. (16)]{goldsmith2005wireless}
\begin{equation}\label{p_e}
P_e(P_r)\approx 2Q\left(\sqrt{\frac{2|h|^2P_rN}{N_0B}}\sin\left(\frac{\pi}{M}\right)\right),
\end{equation}
which is a function of the radar pulse transmit power $P_r$,
and $Q(\cdot)$ denotes the Gaussian $Q$-function.

Next, we analyze the rate performance of dedicated communication signal transmission, i.e., $y_{c,k}[l],N\le l\le N+J-1$.
From \eqref{y_c_k}, it can be obtained that the equivalent input-output relationship for dedicated communication is the AWGN channel as
\begin{equation}
Y_c = h\sqrt{P_cT_c}S + Z,
\end{equation}
where $S$ and $Z$ denote the independent complex Gaussian signalling and noise, respectively, with $S\sim\mathcal{CN}(0,1)$ and $Z\sim~\mathcal{CN}(0,N_0)$.
Therefore, the achievable spectrum efficiency for dedicated communication transmission is \cite{goldsmith2005wireless}
\begin{equation}\label{rate}
\begin{aligned}
\mathcal{R}_c(P_c) &= \frac{J}{J+N}\log_2\left(1+\frac{|h|^2P_cT_c}{N_0}\right)
\\
&=(1-\rho)\log_2\left(1+\frac{|h|^2P_c}{N_0B}\right),
\end{aligned}
\end{equation}
which is a function of the communication power $P_c$ and the pre-log factor accounts for the percentage of the time occupied by the dedicated communication signals.

\section{Numerical Results and Discussions}\label{numerical_results}
\begin{table}[]
\setlength{\abovecaptionskip}{-0.1cm}
\setlength{\belowcaptionskip}{-0.3cm}
\centering
  \caption{Parameter setting}\label{P_set}
\begin{tabular}{|l|l|}
\hline
 Parameter& Value \\ \hline
 Probability of false alarm& $P_{FA}=10^{-8}$ \\ \hline
 Bandwidth& $B=100\ \mathrm{MHz}$  \\ \hline
 Range resolution& $\triangle R =\frac{c}{2B}= 1.5\ \mathrm{m}$  \\ \hline
 PRI&  $T = 10\ \upmu s$\\ \hline
 Maximum unambiguity range&  $R_{\mathrm{ua}} = \frac{c(T-T_p)}{2}=1350\ \mathrm{m}$\\ \hline
 Duty cycle& $\rho=10\ \%$ \\ \hline
 Pulse duration& $T_p=1\ \upmu s$  \\ \hline
 Time-bandwidth product& $N=100$  \\ \hline
 Number of PRIs per CPI& $K=100$  \\ \hline
 CPI& $KT = 1\ \mathrm{ms}$  \\ \hline
 Carrier frequency& $f_c=3.5\ \mathrm{GHz}$  \\ \hline
 FD-ISAC node TX \& RX antenna gain& $G_t=G_r=17\ \mathrm{dBi}$ \\ \hline
 Maximum transmit power & $P_{\max}=1\ \mathrm{W}$ \\ \hline
 Average transmit power & $\bar{P}=0.1\ \mathrm{W}$ \\ \hline
 Commun. RX antenna gain & $G_c = 0 \ \mathrm{dBi}$  \\ \hline
 Noise PSD& $N_0 = -169 \ \mathrm{dBmW/Hz}$ \\ \hline
 Communication distance& $R_{\mathrm{com}} = 400$ m  \\ \hline
 Path-loss exponent for communication link& $\gamma = 2.7$  \\ \hline
 Embedded PSK modulation order& $M = 128$  \\ \hline
 SI channel gain& $|\beta|^2 = -20\ \mathrm{dB}$  \\ \hline
 RCS & $\sigma = 1 \  \mathrm{m^2}$\\ \hline
\end{tabular}
\vspace{-0.4cm}
\end{table}
In this section, we provide simulation results to compare the performance of our proposed FD-ISAC with the conventional HD pulsed radar and communication-centric waveform, and give some insights on monostatic ISAC system design.
The parameter setting is given in Table~\ref{P_set}.
For radar sensing, we assume a line-of-sight (LoS) path between the FD-ISAC node and the target, and the two-way channel gain is usually modelled as $\left|\alpha\right|^2=\frac{G_tG_r\lambda^2\sigma}{(4\pi)^3R^4}$ \cite{richards2010principles}, where $G_t$ and $G_r$ denote the transmit and receive antenna gain of the FD-ISAC node, and $\sigma$ is the {\it radar cross section} (RCS) of the target.
On the other hand, the communication channel gain between the FD-ISAC node and the communication receiver is modelled as $\left|h\right|^2=\frac{G_tG_c\lambda^2}{\left(4\pi\right)^2R_{\mathrm{com}}^\gamma}$ \cite{goldsmith2005wireless}, where $G_c$ denotes the receive antenna gain of the communication receiver, $R_{\mathrm{com}}=400$ m is assumed as the communication distance, and $\gamma=2.7$ denotes the path-loss exponent of the communication channel.

\begin{figure} 
  \centering
  \includegraphics[width=0.45\textwidth]{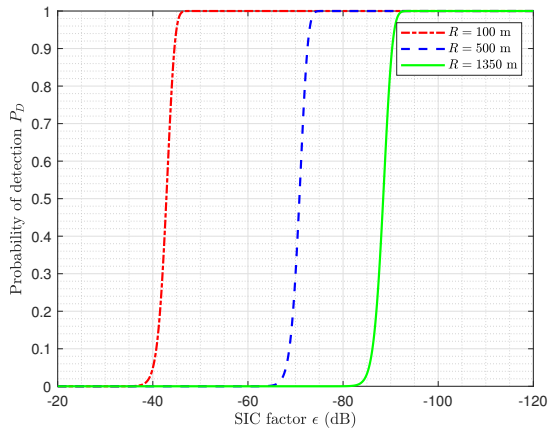}
  \caption{The probability of detection $P_D$ as a function of the SIC factor $\epsilon$ for different target distances $R=100$, $500$, and $1350$ m, where $P_r=P_c=P_{\max}=1 \ \mathrm{W}$.}\label{sic_factor}
\end{figure}

In Fig.~\ref{sic_factor}, we study the probability of detection $P_D$ in \eqref{p_d} versus the SIC factor $\epsilon$ for different target distances $R$, with $P_r=P_c=P_{\max}$, where $P_{\max}=1$ W denotes the maximum transmit power.
It is observed from Fig.~\ref{sic_factor} that as $R$ increases, more powerful SIC capability is required to maintain the desired sensing performance.
For example, as the target distance $R$ increases from $100$ m to $1350$ m, the required SIC increases from $45$ dB to $92$ dB to guarantee the probability of detection $P_D\ge 99\%$.
Such an observation is expected, since the further the target is, the weaker the target echoes is, and hence the more powerful SIC capability is required.

\begin{figure} 
  \centering
  \includegraphics[width=0.48\textwidth]{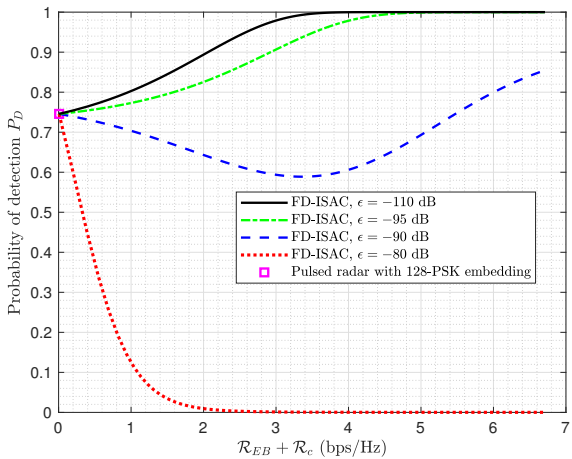}
  \caption{Probability of detection versus communication spectrum efficiency.}\label{Pdrate}\vspace{-0.6cm}
\end{figure}
In Fig.~\ref{Pdrate}, we study the probability of detection $P_D$ versus the total communication spectrum efficiency $\mathcal{R}_t$ for the proposed FD-ISAC scheme with different SIC factors $\epsilon$, where total communication spectrum efficiency $\mathcal{R}_t$ includes the contributions from both PSK embedding with rate $\mathcal{R}_{EB}$ and dedicated communication transmission with rate $\mathcal{R}_c$.
As a benchmark comparison, the HD pulsed radar with PSK  information embedding is also given in the figure.
The target distance is assumed to be $R=R_{\mathrm{ua}}=1350$ m.
The transmit power of the radar pulse is fixed to $P_r=P_{\max}=1\ \mathrm{W}$, and $P_c$ varies from $0$ to $P_{\max}$ so as to achieve different dedicated communication rate $\mathcal{R}_c$.
It is observed from Fig.~\ref{Pdrate} that for the  HD pulsed radar with $128$-PSK information embedding, the probability of detection is $P_D=78.6\%$, and the spectrum efficiency $\mathcal{R}_{EB}$ is only $0.007$ bps/Hz, with the probability of symbol error $P_e\approx 10^{-7}$.
By contrast,
the proposed FD-ISAC scheme can drastically improve the communication spectrum efficiency by increasing the power $P_c$ of the dedicated communication signals, and its impact on the probability of detection $P_D$ is dependent on the SIC capability $\epsilon$.
With weak SIC capabilities, say $\epsilon=-80$ dB, $P_D$ degrades as the communication rate increases, i.e., there is a clear trade-off between sensing and communication performance.
On the other hand, as the SIC capability enhances, e.g., $\epsilon=-95$ or $110$ dB, the probability of detection actually improves with the communication spectrum efficiency.
Such observations are expected, since the dedicated communication signal causes the detrimental SI for radar echoes on one hand, and is also utilized as the additional beneficial signal source for sensing energy accumulation on the other hand (as can be seen from the MF in \eqref{mfer}).
Therefore, as $P_c$ increases, its detrimental impact on $P_D$ dominates if SIC capability is poor, while the reverse is true when the SIC capability is powerful enough.
It is also noted that with the moderate SIC capability, say $\epsilon=-90$ dB, $P_D$ first decreases and then increases with the communication spectrum efficiency, due to the sophisticated impact of $P_c$ on the radar SINR, as shown in \eqref{sinr_1_final}.

\begin{figure} 
  \centering
  \includegraphics[width=0.48\textwidth]{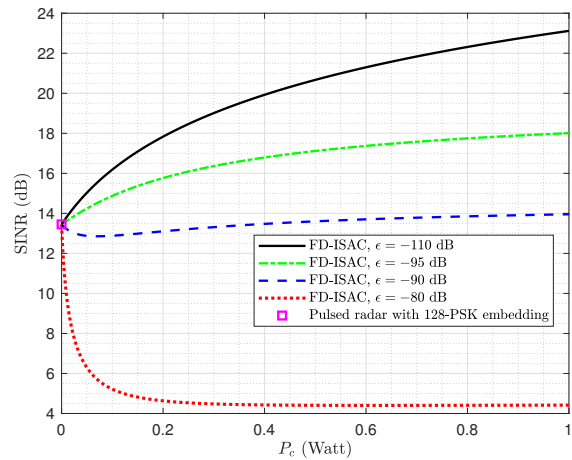}
  \caption{The radar sensing SINR versus the transmit power $P_c$.}\label{Fig2b}\vspace{-0.6cm}
\end{figure}

As a further illustration, Fig.~\ref{Fig2b} shows the SINR of the radar echoes as a function of $P_c$ for different SIC factors $\epsilon$.
It is observed that the results in Fig.~\ref{Fig2b} corroborate those in Fig.~\ref{Pdrate}, which demonstrate that as long as the FD-ISAC node has sufficiently powerful SIC capability, the proposed FD-ISAC scheme is able to achieve mutual benefits for both radar sensing and wireless communication.

\begin{figure} 
  \centering
  \includegraphics[width=0.48\textwidth]{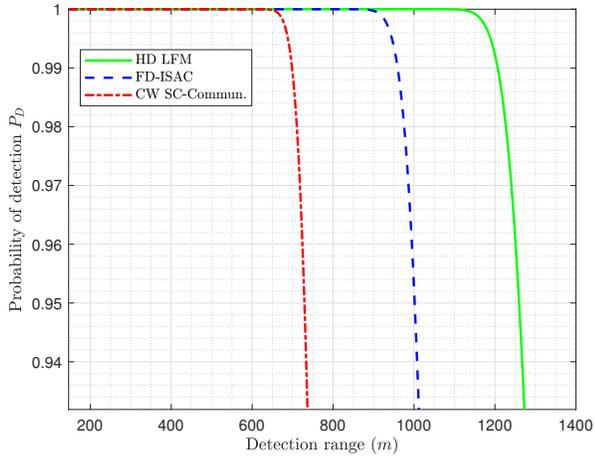}
  \caption{The probability of detection $P_D$ as a function of the detection range for different transmit waveform, i.e., HD pulsed LFM, the proposed FD-ISAC, and SC-Commun waveforms, with $\rho P_r + (1-\rho)P_c=\bar{P}$ and $P_r,P_c\le P_{\max}$, where $\rho=0.1$, $\bar{P}=0.1$ W, $P_{\max}=1$ W, $P_c=0.01$ W and $P_r=0.91$~W. The SIC factor is $\epsilon=-80$ dB.}\label{Pd_SINR}\vspace{-10pt}
\end{figure}
In Fig.~\ref{Pd_SINR}, we study the probability of detection $P_D$ in \eqref{p_d} as a function of the detection range, for different transmit waveforms, i.e., HD pulsed LFM, the proposed FD-ISAC, and CW single-carrier communication-centric (SC-Commun) waveforms.
For fair comparison, we consider both the maximum and average transmit power constraints, i.e., $P_r\le P_{\max}$, $P_c\le P_{\max}$, and $\rho P_r + (1-\rho)P_c=\bar{P}$, where we assume that $\bar{P}=0.1$ W.
The SIC factor is $\epsilon=-80$ dB.
It is observed from Fig.~\ref{Pd_SINR} that compared to the pulsed radar, the maximum detection range of the CW SC-Commun waveform is severely limited due to the serious SI problem.
By contrast, our proposed FD-ISAC scheme can flexibly control the transmit power to mitigate the SI and extend the detection range via reducing $P_c$ while increasing $P_r$.
Specifically, for the HD LFM waveform, the maximum detection range is $1200$ m with $P_D\ge 99\%$ , while that for CW SC-Commun is only $700$ m.
For the proposed FD-ISAC with both peak and average transmit power constraint, by letting $P_c = 0.01$ W and $P_r = 0.91$~W, the detection range is significantly increased to $950$ m for $P_D\ge 99\%$.

\begin{figure} 
  \centering
  \includegraphics[width=0.48\textwidth]{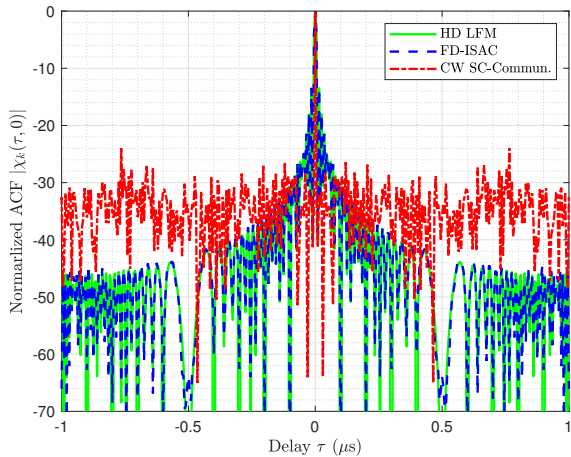}
  \caption{Comparison the ACF for HD LFM, the proposed FD-ISAC and CW SC-Commun, with of $\rho=0.1$, $\bar{P}=0.1$ W, $P_{\max}=1$ W, $P_c=0.01$ W and $P_r=0.91$~W.}\label{AF}\vspace{-10pt}
\end{figure}
In Fig.~\ref{AF}, we compare the normalized ACF in \eqref{ACF-N} for the HD LFM, the proposed FD-ISAC, and CW SC-Commun, to evaluate their sensing resolution capability for range estimation.
It is observed from Fig.~\ref{AF} that the CW SC-Commun waveform leads to ACF that has random PSL, since its ACF is critically dependent on the random communication symbols.
As a consequence, the high instantaneous PSL may occasionally mask the matched filer output of the received signal from the target, rendering the poor sensing performance.
In essence, such non-repeating CW SC-Commun radar can be viewed as noise radar \cite{blunt2016overview}, and if the signal bandwidth is wide enough, the PSL for such noise-like radars degrades as $1/\sqrt{T_{cpi}}$, where $T_{cpi}$ denotes the CPI.
However, such methods require to apply the MF over the complete CPI instead of over each PRI only, rendering high computational and caching complexity.
Furthermore, such random communication signals suffers from high PAPR issue in a data-dependent way, which makes them only suitable for low-power and short-range sensing applications \cite{blunt2016overview}.
On the other hand, the HD pulsed LFM has good and constant ACF for range sensing, where the maximum PSL is $13.2$ dB when the time-bandwidth product is $N=100$, which can be further suppressed by proper windowing, but with the cost of enlarged the mainlobe \cite{blunt2016overview}.
Compared with CW SC-Commun, the proposed FD-ISAC can flexibly control the transmit power to obtain the desired ACF for better range resolution.
Specifically, as shown in Fig.~\ref{AF}, when $P_r=0.91$~W and $P_c=0.01$ W, the normalized ACF of our proposed FD-ISAC almost matches with the HD LFM, which implies good sensing resolution capability.

\section{Conclusion}\label{conclusion}
In this paper, we proposed a novel FD waveform design for monostatic ISAC systems, where the classic pulsed radar waveform is flexibly time-multiplexed with dedicated communication signals.
The probability of detection, AF, and communication spectrum efficiency of the proposed FD-ISAC waveform were derived, and extensive numerical results were provided to compare the performance of the proposed scheme with various benchmark waveforms.
Compared to the existing monostatic schemes that rely on pulsed transmission with information symbol embedding, the proposed FD-ISAC waveform can significantly improve the communication rate and also enhance the probability of target detection, as long as the SI is effectively suppressed.
On the other hand, compared to the communication-centric waveform with random autocorrelation property and degraded sensing performance, the proposed FD-ISAC has better autocorrelation property since it preserves the classic radar waveform for sensing.
In addition, thanks to the FD operation, the proposed FD-ISAC can also mitigate the eclipsing and blind range issues suffered by the classic pulsed radar waveforms, which makes it especially promising for ISAC applications with both distant and  nearby targets.

\begin{appendices}
\section{Proof of \eqref{sid}}\label{appendix a}
To get $x_k^d[l]=\left<x_k(t),\psi(t-lT_c)\right>$ in \eqref{sid}, we consider following two cases separately:

\emph{Case-I:} $0\le t \le T_p$, then
\begin{equation}\label{a-case1}
\begin{aligned}
&\left<x_k(t),\psi(t-lT_c)\right> \\
&=\left<\sqrt{P_r}\omega_k p(t),\psi(t-lT_c)\right> \\
&=\int_{0}^{T_p}\sqrt{P_r}\omega_k p(t)\psi^*(t-lT_c)\mathrm{d}t \\
&=\int_{0}^{T_p}\sqrt{P_rT_c}\omega_k\sum\limits_{n=0}^{N-1}c[n]\psi(t-nT_c)\psi^*(t-lT_c)\mathrm{d}t \\
&=\sqrt{P_rT_c}\omega_kc[l], 0\le l \le N-1,
\end{aligned}
\end{equation}
where we have used the expression for $p(t)$ in \eqref{p_t} and that the autocorrelation function of $\psi(t)$ is $R_{\psi}(t)=\delta(t)$.

\emph{Case-II:} $T_p<t\le T$, then
\begin{align}\label{a-case2}
&\left<x_k(t),\psi(t-lT_c)\right> \notag\\
&=\left<\sqrt{P_c}g_k(t-T_p),\psi(t-lT_c)\right> \notag\\
&=\int_{T_p}^{T}\sqrt{P_c}g_k(t-NT_c)\psi^*(t-lT_c)\mathrm{d}t \notag\\
&=\int_{T_p}^{T}\sqrt{P_cT_c}\sum\limits_{j=0}^{J-1} s_k[j]\psi(t-(N+j)T_c)\psi^*(t-lT_c)\mathrm{d}t \notag\\
&=\sqrt{P_cT_c}s_k[l-N], N\le l \le N+J-1.
\end{align}
Therefore, by combining \eqref{a-case1} and \eqref{a-case2}, we get \eqref{sid}.

\section{Proof of \eqref{project x_k}}\label{appendix b}
To prove \eqref{project x_k} for getting $\left<x_k(t-\tau),\psi(t-lT_c)\right>$ based on \eqref{x_k_tau}, we need to consider the following three cases:

\emph{Case-I:} $0\le t\le \tau$, then
\begin{align}\label{caseI}
&\left<x_k(t-\tau),\psi(t-lT_c)\right> \notag\\
&=\left<\sqrt{P_c}g_{k-1}(t+T-\tau-T_p),\psi(t-lT_c)\right> \notag\\
&=\left<\sqrt{P_c}g_{k-1}(t+(J-n_\tau)T_c),\psi(t-lT_c)\right> \notag\\
&=\int_{T_p}^{\tau}\sqrt{P_c}g_{k-1}(t+(J-n_\tau)T_c)\psi^*(t-lT_c)\mathrm{d}t \notag\\
&=\int_{T_P}^{\tau}\sqrt{P_cT_c}\sum\limits_{j=0}^{J-1}s_{k-1}[j]\psi(t+(J-n_\tau-j)T_c)\psi^*(t-lT_c)\mathrm{d}t
\notag\\
&=\sqrt{P_cT_c}s_{k-1}[J-n_\tau+l], 0\le l \le n_\tau-1.
\end{align}

\emph{Case-II:} $\tau<t\le \tau+T_p$, then
\begin{align}\label{caseII}
&\left<x_k(t-\tau),\psi(t-lT_c)\right> \notag\\
&=\left<\sqrt{P_r}\omega_k p(t-\tau),\psi(t-lT_c)\right> \notag\\
&=\sqrt{P_rT_c}\omega_k\int\limits_{\tau}^{\tau+T_p}p(t-\tau)\psi^*(t-lT_c)\mathrm{d}t \notag\\
&=\sqrt{P_rT_c}\omega_k\int\limits_{\tau}^{\tau+T_p}\sum\limits_{n=0}^{N-1}c[n]\psi(t-(n_\tau+n)T_c)\psi^*(t-lT_c)\mathrm{d}t\notag\\
&=\sqrt{P_rT_c}\omega_k c[l-n_\tau], n_\tau\le l \le N+n_\tau - 1.
\end{align}

\emph{Case-III:} $\tau+T_p< t\le T$, then
\begin{align}\label{caseIII}
&\left<x_k(t-\tau),\psi(t-lT_c)\right> \notag\\
&=\left<\sqrt{P_c}g_k(t-\tau-T_p),\psi(t-lT_c)\right> \notag\\
&=\left<\sqrt{P_c}g_k(t-(n_\tau+N)T_c),\psi(t-lT_c)\right> \notag\\
&=\int\limits_{\tau+T_p}^{T}\sqrt{P_c}g_k(t-(n_\tau+N)T_c)\psi^*(t-lT)c)\mathrm{d}t \notag\\
&=\int\limits_{\tau+T_p}^{T}\sqrt{P_cT_c}\sum\limits_{j=0}^{J-1}s_k[j]\psi(t-(N+\tau+j)T_c)\psi^*(t-lT_c)\mathrm{d}t\notag\\
&=\sqrt{P_cT_c}s_k[l-N-n_\tau], N+n_\tau\le l \le N+J-1.
\end{align}
Therefore, by combining \eqref{caseI}, \eqref{caseII}, and \eqref{caseIII}, we get \eqref{project x_k}.

\end{appendices}

\bibliographystyle{IEEEtran}
\bibliography{RadarRef,SIC_ref,Com_ref,ISAC}

\end{document}